\documentclass[reprint,prb,superscriptaddress]{revtex4-2}

\usepackage{xcolor}
\usepackage{graphicx}
\usepackage{ulem}
\usepackage{float}
\usepackage{amsmath,amssymb}
\newcommand{\up}{\uparrow}
\newcommand{\dn}{\downarrow}
\newcommand{\kv}{\mathbf{k}}
\newcommand{\rv}{\mathbf{r}}
\newcommand{\Rv}{\mathbf{R}}

\newcommand{\qv}{\mathbf{q}}
\renewcommand{\Im}{\operatorname{Im}}
\renewcommand{\Re}{\operatorname{Re}}
\newcommand{\abs}[1]{\left|#1\right|}
\newcommand{\av}[1]{\left\langle#1\right\rangle}
\newcommand{\avv}[1]{\left\langle\!\left\langle#1\right\rangle\!\right\rangle}
  
\usepackage{comment}

\newcommand{\conf}{\mathcal{F}}

\DeclareMathOperator{\FT}{FT}
\DeclareMathOperator{\Tr}{Tr}

\begin{document}
\title{Benchmarking the ladder Dual Fermion approach for the Falicov-Kimball model}

\author{Akshat Mishra}
\affiliation{NanoLund and Division of Mathematical Physics,
    Department of Physics,
    Lund University, Lund,
    Sweden}

\author{Hugo U. R. Strand}
\affiliation{School of Science and Technology, Örebro University, SE-701 82 Örebro, Sweden}

\author{Erik G. C. P. van Loon}
\affiliation{NanoLund and Division of Mathematical Physics,
    Department of Physics,
    Lund University, Lund,
    Sweden}
\affiliation{LINXS Institute of advanced Neutron and X-ray Science (LINXS)}    

\begin{abstract}
Strong electronic correlations generally require non-perturbative treatment. Local correlations are captured by dynamical mean-field theory while nonlocal correlations can be treated with diagrammatic extensions such as the Dual Fermion approach. Dual Fermion is built on physically motivated, but in principle uncontrolled approximations, so careful benchmarking is needed to understand the strengths and limitations of the method.
In this work, we benchmark ladder Dual Fermion and dynamical mean-field theory for the Falicov-Kimball model with the exact classical Monte Carlo solution. We focus on the thermodynamics, electronic structure and susceptibility, especially at the combined frequency and momentum structure, and find that Dual Fermion clearly outperforms dynamical mean-field theory.
Somewhat surprisingly, Dual Fermion is not as accurate for the relation between orbital density versus chemical potential in the doped system. 
These results demonstrate the need for rigorous benchmarking of diagrammatic extensions of dynamical mean-field theory for models with inequivalent orbitals, which is essential for modelling materials.
\end{abstract}
    
\maketitle

\section{Introduction}

Electronic correlations in quantum materials drive a multitude of phases and collective phenomena such as superconductivity~\cite{dagotto1994correlated}, metal-insulator transitions~\cite{mott1968metal}, and charge-density waves~\cite{Hansmann13,xu2024coexistence}, as observed both in real materials~\cite{basov2017towards}, quantum simulators~\cite{esslinger2010fermi}, and theoretical models~\cite{qin2022hubbard,arovas2022hubbard}.
The correlations are governed by the competition between the electronic interaction and the electronic hopping, which we denote by $U$ and $t$, respectively.
Theoretically, the weak ($U/t \ll 1$) and strong coupling ($U/t \gg 1$) regimes are amenable to perturbation expansions. However, the intermediate regime ($U/t \sim 1$) is particularly challenging due to the lack of a small parameter, and can only be studied using non-perturbative computational methods.

Here, dynamical mean-field theory (DMFT)~\cite{GeorgesRMP1996} is a state-of-the-art method for non-perturbative treatment of strong local correlations. While formally exact in infinite dimensions \cite{Metzner89}, DMFT amounts to a momentum-independent self-energy approximation in finite dimensions. Thus, while DMFT excels at describing local correlations like the Mott transition, it fails spectacularly in regimes with strong non-local correlation by, e.g., violating the Hohenberg-Mermin-Wagner theorem \cite{Hohenberg67,MerminWagner}.

To remedy this, several non-local extensions of DMFT, from cluster approaches~\cite{Maier05}, like the DCA \cite{HettlerPRB1998, HettlerPRB2000}, CDMFT \cite{KotliarPRL2001}, nested clusters~\cite{PhysRevLett.75.113, PhysRevB.69.205108, PhysRevB.97.125141}, to diagrammatic approaches like the Dual Fermion (DF)~\cite{RubtsovPRB2008}, Dual Boson~\cite{rubtsov2012dual}, D$\Gamma$A~\cite{Toschi07}, EDMFT~\cite{Sengupta95,Si96}, EDMFT+$GW$~\cite{Sun02}, fRG~\cite{RevModPhys.84.299,Taranto14}, TRILEX \cite{AyralPRB2015,AyralPRB2016,AyralPRL2017}, QUADRILEX \cite{Quadrilex2016}, Dual-TRILEX \cite{StepanovPRB2019} and single-boson exchange~\cite{Krien19}, have been developed.
However, all of these methods contain approximations that are physically motivated but generally uncontrolled (especially close to phase transitions~\cite{Schafer15}).
To address this situation extensive benchmarking has been performed on the two-dimensional Hubbard model, comparing both approximate methods and statistically exact methods in controlled regimes~\cite{LeBlanc15,Schafer15,Schafer21}.
The benchmarks have largely been limited to thermodynamical quantities and single-particle spectra, due to the challenge of obtaining frequency-momentum-resolved results, in particular for the susceptibility.

The Falicov-Kimball model~\cite{FalicovKimball,Freericks03} provides a good opportunity for benchmarking beyond-DMFT methods. It is one of the simplest canonical lattice models for correlated electrons with a rich phase diagram exhibiting disordered-gapless, insulating and charge-ordered phases. Although the model is mostly of theoretical interest, there have been proposals to implement it using ultracold atoms in optical lattices~\cite{Maska08}. The reason for the relative simplicity of the Falicov-Kimball model is the fact that half the electrons in the model are effectively classical. In infinite dimensions, DMFT for the Falicov-Kimball model is not only exact but also analytically solvable~\cite{Freericks03}. In finite dimensions, the lattice problem can be effectively reduced into an ensemble of non-interacting problems, that can be sampled using classical Monte Carlo~\cite{Maska06,AntipovPRL2016} instead of quantum Monte Carlo~\footnote{Similarly, exact enumeration or classical Monte Carlo can be used as the impurity solver in cluster DMFT for the Falicov-Kimball model~\cite{HettlerPRB1998,HettlerPRB2000}, instead of the quantum Monte Carlo that is used for the Hubbard model.}.
While diagrammatic extensions to DMFT previously have been applied to the Falicov-Kimball model to determine critical exponents~\cite{AntipovPRL2014}, an in-depth benchmarking of the spatial and temporal structure of the single-particle response and susceptibility is lacking.

In this paper, we benchmark DMFT and ladder DF against classical Monte Carlo performed for the Falicov-Kimball model on the square lattice, both at half-filling and away from half-filling. We choose the ladder approximation~\cite{Hafermann09} in DF as it efficiently describes the dominant spatial fluctuations and is much simpler than variants summing additional diagrams~\cite{Iskakov16,Gukelberger17,Astleithner20}. We evaluate single-particle observables such as the Green's function, self-energy and electronic occupation numbers as well as two-particle observables like the susceptibility. We investigate the model in a challenging regime close to the phase transition to a charge density wave (CDW) phase. While similar benchmarks have been carried out for the Hubbard model in the past~\cite{LeBlanc15,Schafer21},
we show the qualitative and quantitative comparison of the full frequency and momentum structure of the Green's function, self-energy and dynamic $cc$-susceptibility against exact Monte Carlo results, in different doping, correlation and temperature regimes. We perform our benchmark for the two-dimensional square lattice because nonlocal fluctuations are stronger in low dimensions, Appendix~\ref{app:3d} shows similar benchmarking in the three-dimensional cubic lattice. This study opens the possibility of exploring the use of DF to study larger and more complex correlated quantum systems in a computationally viable and controlled manner, even close to phase transitions. 

\subsection{The Hubbard model}

The Hamiltonian of the Hubbard model \cite{Hubbard1, GutzwillerPRL1963, KanamoriPTP1963} is
\begin{equation}
  H_\text{Hub} =
  - \sum_{\langle i, j \rangle, \sigma} t_\sigma \hat{c}^\dagger_{\sigma,i} \hat{c}_{\sigma,j}
  + U \sum_{i} \hat{n}_{\uparrow,i} \hat{n}_{\downarrow,i}
  - \sum_{i, \sigma}\mu_\sigma \hat{n}_{\sigma,i}
  \label{eq:Hubbard}
\end{equation}
where $\hat{c}^\dagger_{\sigma,i}$ creates an electron on site $i$ with spin $\sigma \in \{ \uparrow, \downarrow \}$, $\hat{n}_{\sigma,i} \equiv \hat{c}^\dagger_{\sigma,i} \hat{c}^{\phantom{\dagger}}_{\sigma,i}$ is the spin-density operator,
$U$ is the local Hubbard interaction,
while $t_\sigma$ and $\mu_\sigma$ are the spin-dependent nearest-neighbor hopping and chemical potential, respectively.
In the standard formulation the Hubbard model has $SU(2)$ spin symmetry with $t_\uparrow=t_\downarrow$ and $\mu_\uparrow=\mu_\downarrow$.

On the square lattice, the Hubbard model features strong antiferromagnetic correlations~\cite{Schafer15} but for any finite temperature the Hohenberg-Mermin-Wagner theorem~\cite{Hohenberg67,MerminWagner} prohibits spontaneous breaking of the continuous $SU(2)$ spin symmetry. This, in turn, guarantees that no antiferromagnetic ordering can occur.

The presence of antiferromagnetic fluctuations with extremely long but finite correlation length is one of the main challenges when describing the 2D Hubbard model in DMFT-based approaches~\cite{Schafer15}, since this kind of approximate theory has a strong tendency to favor phase transitions. Diagrammatically, the fluctuations that should restore the Mermin-Wagner theorem involve very high-order diagrammatic processes and self-consistent feedback between single-particle and two-particle Green's functions in the diagrammatic theory~\cite{Schafer15,vanLoon18}.

\subsection{The Falicov-Kimball model}

The Falicov-Kimball model can be obtained from the Hubbard model in Eq.~\eqref{eq:Hubbard} by mapping the two spin states ($\uparrow$ and $\downarrow$) to one flavor of mobile $c$-electrons with finite hopping ($t_c = t_\uparrow > 0$) and a second flavor of immobile $f$-electrons with zero hopping ($t_f = t_\downarrow = 0$).
In other words, the $SU(2)$ symmetry of the two kinds of fermions is broken explicitly via the hopping.

Thus, the Hamiltonian of the Falicov-Kimball model takes the form
\begin{multline}
  H_{\text{FK}} = - t_c \sum_{\langle i, j \rangle} \hat{c}^\dagger_{i} \hat{c}_{j}
  + U \sum_{i} \hat{n}_{c,i} \hat{n}_{f,i} \\
 -\mu_c \sum_{i} \hat{n}_{c,i}
  - \mu_f \sum_{i} \hat{n}_{f,i} \label{eq:FK}
\end{multline}
where $\hat{c}^\dagger_{i}$ and $\hat{f}^\dagger_{i}$ are the creation operators for $c$ and $f$-electrons on site $i$, respectively, $\hat{c}_{i}$ and $\hat{f}_{i}$ the corresponding annihilation operators, and $\hat{n}_{i,c}=\hat{c}^\dagger_i \hat{c}^{\phantom{\dagger}}_i$ and $\hat{n}_{i,f}=\hat{f}^\dagger_i \hat{f}^{\phantom{\dagger}}_i$ the number operators.

The antiferromagnetic correlations in the Hubbard model turn into $c$-$f$ checkerboard order in the Falicov-Kimball model~\cite{Freericks03}. This checkerboard order is often called a charge-density wave (CDW), although it should be noted that the total charge density is uniform and the order is frozen because the $f$ electrons are immobile. The Mermin-Wagner theorem is not applicable in the Falicov-Kimball model, since the symmetry that is broken is discrete instead of continuous, so the CDW has a finite critical temperature, $T_c$. 

The fact that there is a genuine phase transition at finite temperature means that there is a diverging correlation length at $T_c$, which is visible in DMFT and its extensions as a divergence in the susceptibility~\cite{AntipovPRL2014}. With the single-site impurity model used here and in Ref.~\cite{AntipovPRL2014}, it is not possible to enter into the ordered phase, so this phase is excluded from our benchmark.

On the square lattice studied here, both the Hubbard and Falicov-Kimball models are particle-hole symmetric at $\mu=U/2$ which guarantees half-filling for both electron flavors. However, the models behave differently away from half-filling. For the $SU(2)$ symmetric Hubbard model, $\mu_\uparrow=\mu_\downarrow$ gives $\langle \hat{n}_\up \rangle = \langle \hat{n}_\dn \rangle$, unless there is spontaneous symmetry breaking. For the Falicov-Kimball model, on the other hand, $\langle \hat{n}_f \rangle \neq \langle \hat{n}_c \rangle$ even when $\mu_f=\mu_c$. In fact, this is even true in the non-interacting Falicov-Kimball model. The reason for this difference is that the $c$-electrons have dispersive bands while the $f$-electrons are immobile and form a flat band. 


\section{Falicov-Kimball Model - Basics and Monte Carlo}

The Falicov-Kimball model comprises two different species of spinless electrons: the fixed $f$-electrons and the itinerant $c$-electrons with a hopping energy, $t$, and an onsite Coulomb interaction $U$. The Hamiltonian can be written as in Eq.~(\ref{eq:FK}), with $\mu_c = \mu_f = \mu$ in our case.

In the grand canonical ensemble, the partition function $Z$ at inverse temperature $\beta$ is
\begin{align}
  Z &= \Tr \exp(-\beta H_\text{FK})
  \\
  &= \Tr_f \Tr_c \exp(-\beta H_c[\hat{c}^\dagger,\hat{c};\hat{n}_{f,i}]) \exp(\beta \mu_f \sum_i \hat{n}_{f,i}), \nonumber
\end{align}
where the trace over the electronic degrees of freedom has been split into separate traces over $f$ and $c$-electrons and the $c$-electron Hamiltonian is
\begin{align}
  H_c[\hat{n}_{f,i}] &=
  - t_c \sum_{\langle i, j \rangle} \hat{c}^\dagger_{i} \hat{c}_{j}
  + U \sum_{i} \hat{n}_{c,i} \hat{n}_{f,i} 
 -\mu_c \sum_{i} \hat{n}_{c,i} \nonumber.
\end{align}
The $f$-electrons are entirely classical in the Falicov-Kimball model, in the sense that all $\hat{n}_{f,i}$ commute with the Hamiltonian and are therefore good quantum numbers. Thus, the $f$-trace can be replaced by a sum over $f$-electron configurations $\conf$ giving the partition function
\begin{equation}  
  Z = \sum_\conf \exp(\beta \mu_f N_f^\conf) \Tr_c \exp(-\beta
  H_c^\conf)
  \, , \label{eq:partition:1}
\end{equation}
where $H^\conf_c \equiv H_c[\langle \hat{n}_{f,i} \rangle_\conf]$ and $N_f^\conf \equiv \sum_i \av{\hat{n}_{f,i}}_\conf$ is the total number of $f$-electrons in the configuration $\conf$. In the sum, every configuration $\conf$ can be represented by a string of bits, indicating whether site $i$ contains an $f$ electron or not, i.e.\ $\av{\hat{n}_{f,i}}_\conf \in \{0, 1\}$. 

The central property of the Falicov-Kimball model that emerges from Eq.~\eqref{eq:partition:1} is that the $f$-electrons are classical while the $c$-electrons are quantum but by themselves non-interacting. By integrating out the quantum degrees of freedom, one ends up with an effective action for the classical degrees of freedom, which can be treated using classical Monte Carlo. Explicitly,
\begin{align}
 \Omega(\conf) &\equiv -\mu_f N_f^\conf - \frac{1}{\beta} \ln\left( \Tr_c \exp(-\beta H_c^\conf) \right), \label{eq:freeenergy} \\
 Z &= \sum_\conf \exp\left(-\beta \Omega(\conf)\right).
\end{align}
Here, $\Omega(\conf)$ is the free energy of a configuration of $f$-electrons, which includes the chemical potentials for both flavors, the interaction energy, the kinetic energy of the $c$-electrons and the entropy of the $c$-electrons. 

To perform Metropolis Monte Carlo, the acceptance ratio for two arbitrary $f$-configurations $\conf_1$, $\conf_2$ is needed. Here, in accordance with Eq.~\eqref{eq:freeenergy}, this ratio is
\begin{align}
 \frac{p(\conf_1)}{p(\conf_2)} = \exp\left(\beta \Omega(\conf_2)- \beta \Omega(\conf_1) \right). \label{eq:acceptance}
\end{align}
It should be noted that $\Omega(\conf)$ also implicitly depends on $\beta$ through Eq.~\eqref{eq:freeenergy}.
Apart from this, the Monte Carlo procedure is essentially similar to Metropolis Monte Carlo for the Ising model~\cite{newman1999monte}.
The Markov chain moves used in the simulations are local $f$-electron occupation changes (applying a boolean \texttt{not} on the corresponding bit) and $f$-electron pair exchange of randomly selected pairs of one filled and one empty site.
Similarly, observables that relate to the $f$-electrons only, such as the average number of $f$-electrons and their spatial correlation, are determined by averaging over $f$-configurations in the Monte Carlo procedure just like in the Ising model, since they are completely classical. 

The situation is slightly different for observables and Green's functions related to the $c$-electrons. The local number operator $\hat{n}_{c,i}$ does not commute with the Hamiltonian, which shows that the $c$-electrons are governed by (quantum) dynamics. From the point of view of the $c$-electrons, every $f$-electron configuration $\conf$ corresponds to a non-uniform potential energy landscape and the averaging over $f$-electrons can therefore be seen as averaging over disorder configurations. In the following subsections, we derive the single-particle and two-particle Green's function of the $c$-electrons to illustrate this point. 

\subsection{Single-particle observables}

We start with the single-particle Green's function for the $c$-electrons. We consider a finite system of size $N$, so that the tight-binding Hamiltonian $\hat{H}_0$ without any $f$-electrons present is the $N\times N$ matrix with nonzero elements,
\begin{align}
\left(\hat{H}_0\right)_{i,j} = \begin{cases}
                                 -\mu_c, &\text{if $i=j$} \\
                                 -t, &\text{if $i$, $j$ neighbors}                                  \end{cases}. 
\end{align}
Note that the chemical potential $\mu_c$ is included in $\hat{H}_0$ for notational convenience. The corresponding bare Green's function is 
\begin{align}
\hat{\mathcal{G}}_0(\nu_n) = \left(i\nu_n \hat{I} - \hat{H}_0\right)^{-1}. \label{eq:g0}
\end{align}
Here, $\nu_n$ is the $n^\text{th}$ fermionic Matsubara frequency, $\hat{I}$ is the identity matrix and the inverse is a matrix inverse. 

An $f$-electron configuration $\conf$ gives a diagonal matrix with elements (on-site potentials) 
 \begin{align}
	\left(\hat{H}_\conf\right)_{i,i} = U \av{\hat{n}_{f,i}}_\conf \, .
 \end{align}
The corresponding $c$-electron Green's function for this $f$-electron configuration $\conf$ is 
\begin{align}
  \hat{G}_\conf(\nu_n) &= \left( i\nu_n \hat{I} -\hat{H}_0 - \hat{H}_\conf \right)^{-1} . \label{eq:gf:def}
\end{align}
Since $\left(\hat{G}_\conf(\nu_n) \right)^{-1} = \left(\hat{\mathcal{G}}_0(\nu_n) \right)^{-1} - \hat{H}_\conf$, one can interpret $\hat{\Sigma}_\conf=\hat{H}_\conf$ as the interaction-induced self-energy for this configuration. Note that this self-energy is frequency-independent since it comes from a static potential. 

The calculation of the expectation value, denoted by double brackets, involves a sum over the $f$-electron configurations $\conf$, weighted by their probabilities $p(\conf)$,
\begin{align}
  \avv{\hat{G}(\nu_n)} &= 
   \sum_\conf p(\conf) \hat{G}_{\conf}(\nu_n). \label{eq:gf:averaging}
\end{align}
In the Monte Carlo procedure, the exact probabilities $p(\conf)$ are replaced by a set of $M$ configurations $\{ \conf_\alpha \}_{\alpha=1}^M$, sampled according to the probability distribution $p({\conf})$ using Eq.~\eqref{eq:acceptance} and the Metropolis algorithm~\cite{newman1999monte},
\begin{align}
   \avv{\hat{G}(\nu_n)} &\overset{\text{MC}}{\approx} 
   \frac{1}{M}\sum_{\alpha=1}^M \hat{G}_{\conf_\alpha}(\nu_n), \label{eq:gf:averaging:mc}
\end{align}
which becomes exact as $M \rightarrow \infty$.

In Eqs.~\eqref{eq:gf:averaging} and \eqref{eq:gf:averaging:mc}, the hat on $\hat{G}$ denotes the fact that it is an $N\times N$ matrix in real space, where $N$ is the number of lattice sites. We write $\avv{\hat{G}}_{i,j}$ for the matrix element corresponding to propagation from $j$ to $i$. The individual elements on the right-hand side of Eq.~\eqref{eq:gf:averaging} lack translation symmetry due to the non-uniform potential energy landscape created by the $f$-electrons, but the $\conf$-averaging restores the translation symmetry, so that the matrix $\avv{\hat{G}}_{i,j}$ on the left-hand side depends on the displacement vector $\mathbf{r}_i-\mathbf{r}_j$ only. For the Monte Carlo sampling, the translation symmetry is only reinstated as $M\rightarrow \infty$. In our analysis using finite $M$, we enforce the translation symmetry by defining $G(\nu_n,\Rv)$ (without hat) as 
\begin{align}
 G(\nu_n,\Rv) = \frac{1}{N} \sum_{j} \avv{\hat{G}(\nu_n)}_{\mathbf{r}_j+\Rv,\mathbf{r}_j}.\label{eq:definition:g:final}
\end{align}
Now, since $G(\nu_n,\Rv)$ depends on the displacement vector $\Rv$ only, it can be Fourier transformed to $G(\nu_n,\kv)$ in the usual way. Finally, the self-energy of the Falicov-Kimball model is defined using the Dyson equation as
\begin{align}
 G(\nu_n,\kv) = \mathcal{G}_0(\nu_n,\kv) + \mathcal{G}_0(\nu_n,\kv) \Sigma(\nu_n,\kv) G(\nu_n,\kv), \label{eq:dyson}
\end{align}
where $\mathcal{G}_0(\nu_n,\kv)$ is the Green's function without $f$-electrons as defined in Eq.~\eqref{eq:g0}. Note that $\Sigma(\nu_n,\Rv)\neq \frac{1}{N}\sum_{j} \avv{\hat{\Sigma}_{\mathbf{r}_j+\Rv,\mathbf{r}_j}}$, i.e., the self-energy of the Falicov-Kimball model cannot be interpreted as the $\conf$-average of a per-configuration self-energy.  A clear example of this fact is that all $\hat{\Sigma}_\conf$ are frequency-independent while $\Sigma(\nu_n, \mathbf{k})$ has non-trivial frequency dependence induced by the interaction effects from the $f$-electron configuration averaging.
Phrased differently, the act of configuration averaging does not commute with the inversion necessary to obtain the self-energy from the Dyson equation in Eq.~\eqref{eq:dyson}. 

\subsection{Two-particle observables}

Following a scheme similar to the one above, we can calculate two-particle observables as well. As with $G$ and $\Sigma$, it is necessary to first calculate $f$-configuration-averaged expectation values before performing subtractions and amputations to obtain susceptibilities and vertices. In general, those operations do not commute with the configuration averaging. We start with $G^{(2),\conf}_{\rv\rv\rv'\rv'}(\omega,\nu,\nu')=\av{c^\dagger_\rv(\nu) c_\rv(\nu+\omega) c^\dagger_{\rv'}(\nu'+\omega) c_{\rv'}(\nu')}_\conf$, which is the $c$-electron two-particle Green's function for a specific disorder configuration $\conf$. Here, the PH frequency convention is used and factors of $\beta$ are dropped \cite{TPRF}. 

For any configuration $\conf$, the $c$-electrons themselves are non-interacting, so Wick's theorem can be applied to the expectation values corresponding to each individual configuration $\conf$, 
\begin{align}
 &\avv{G_{\rv\rv\rv'\rv'}^{(2)}(\omega,\nu,\nu')}= \sum_\conf p_\conf G_{\rv\rv\rv'\rv'}^{(2),\conf}(\omega,\nu,\nu') \notag \\
 &=\sum_\conf p_\conf \left[  \delta_{\omega,0} G_{\rv\rv}^{0,\conf}(\nu)G_{\rv'\rv'}^{0,\conf}(\nu')-\delta_{\nu\nu'} G_{\rv'\rv}^{0,\conf}(\nu) G_{\rv\rv'}^{0,\conf}(\nu+\omega) \right].
\end{align}
Here, to later obtain a single-momentum object, we have used only two position coordinates $\rv$, $\rv'$ instead of the general case $\rv_1$, $\rv_2$, $\rv_3$, $\rv_4$.
The last line shows that $\avv{G^{(2)}}$ is a combination of two-frequency objects instead of a completely three-frequency object. This is a specific feature of non-interacting systems like the Falicov-Kimball model~\cite{AntipovPRL2014}. 

To obtain the non-local, single-momentum, single-frequency, dynamical susceptibility $\chi(\omega)$ of the $c$-electrons, in a particular channel, we need to subtract the ``bubble'' of disorder-averaged Green's functions $\avv{\hat{G}}$ in the other channel and sum over the fermionic frequencies $\nu$, $\nu'$ (we follow the notational conventions of TPRF~\cite{TPRF}).
\begin{align}
 \chi^=_{\rv\rv\rv'\rv'}(\omega) &= \sum_{\nu,\nu'} \chi^=_{\rv\rv\rv'\rv'}(\omega,\nu,\nu') \notag \\
 &= \sum_{\nu,\nu'} \avv{G^{(2)}_{\rv\rv\rv'\rv'}(\omega,\nu,\nu')} \notag \\
 &\phantom{=}- \delta_{\omega,0} \sum_\nu G(\nu,\rv-\rv) \sum_{\nu'} G(\nu',\rv'-\rv') \notag \\ 
 &= \delta_{\omega,0} \sum_\conf p_\conf \sum_{\nu,\nu'} G_{\rv\rv}^{0,\conf}(\nu) G_{\rv'\rv'}^{0,\conf}(\nu') \notag \\
 &\phantom{=}- \sum_\conf p_\conf \sum_{\nu\nu'} \delta_{\nu\nu'}  G_{\rv'\rv}^{0,\conf}(\nu) G_{\rv\rv'}^{0,\conf}(\nu+\omega) \notag \\
 &\phantom{=}- \avv{n_{c,\rv} }\avv{n_{c,\rv'}} \delta_{\omega,0} \notag \\
 &= \delta_{\omega,0} \sum_\conf p_\conf n_{c,\rv}^{0,\conf} n_{c,\rv'}^{0,\conf} \notag \\
 &\phantom{=}- \sum_\conf p_\conf \sum_\nu G_{\rv'\rv}^{0,\conf}(\nu) G_{\rv\rv'}^{0,\conf}(\nu+\omega) \notag \\
 &\phantom{=}- \avv{n_{c,\rv} }\avv{n_{c,\rv'}} \delta_{\omega,0} \notag \\
 &= \delta_{\omega,0} \left[ \avv{n_{c,\rv} n_{c,\rv'}}-\avv{n_{c,\rv}}\avv{n_{c,\rv'}} \right] \notag \\
 &\phantom{=}- \frac{1}{\beta} \FT_{\tau\rightarrow \omega} \sum_\conf p_\conf  G^{0,\conf}_{\rv'\rv}(\tau) G^{0, \conf}_{\rv\rv'}(\beta-\tau). \label{eq:chi:qw}
\end{align}
Here, the first term in the final equation is the statistical covariance of the electron density between sites $\rv$ and $\rv'$ due to the averaging over $f$-electron configurations, while the second term contains the Friedel oscillations, i.e., the charge-correlations in non-interacting systems, averaged over disorder configurations. Finally, assuming translational invariance after configuration averaging, $\chi^=$ can be transformed to momentum space, giving $\chi^=(\omega,\mathbf{q})$. See more details of our implementation in Appendix \ref{app:code}. \\

It is also possible to consider a more generalized object, $G^{(2)}_{\rv_1\rv_2\rv_3\rv_4}$, which after averaging and Fourier transforms corresponds to $G^{(2)}(\kv,\qv,\qv')$. From this $G^{(2)}$, it is possible to extract the three-momentum vertex $\Gamma(\kv,\qv,\qv')$, which is the two-particle analog of the self-energy $\Sigma$. The momentum structure of the vertex plays an important role in diagrammatic approximations, as discussed in more detail in the subsequent section.

 
\section{Dual Fermion and DMFT for Falicov-Kimball}

DMFT and DF are approximate methods for interacting electron systems that make it possible to study large systems at a moderate computational cost. Both are based on a single-site impurity model whose bath is determined self-consistently to cheaply represent the most important local correlations in the system. The bath is described by an effective hybridization function, $\Delta_c(\nu_n)$ or equivalently by the so-called Weiss field on the impurity,
\begin{align}
 \mathcal{G}_0(\nu_n) = [i\nu_n + \mu - \Delta_c(\nu_n)]^{-1} \label{eq:weiss}
\end{align}
The Weiss field $\mathcal{G}_0$ is the non-interacting impurity Green's function, $G_{\conf=0}$. Now there are two possible f-electron configurations with the impurity site having an $f$-electron, $\conf = 1$, or not, $\conf = 0$. We can then write the partition function as a combination of these two possibilities \cite{Brandt1989}.

\begin{align}
  Z &= Z_{\conf=0} + Z_{\conf=1}
  ,
\end{align}
with $Z_{\conf} = 2 \exp \left( \beta \mu_f \conf + \sum_n  e^{i\nu_n 0^+} \cdot \ln \frac{\mathcal{G}_0^{-1}(\nu_n) - U \conf}{i \nu_n} \right)$.

Further, similar to Eq.~\eqref{eq:gf:averaging}, the Green's function for the impurity model is obtained by averaging over the $f$-electron configurations occurring with probabilities $p(\conf)$,
\begin{align}
 G_\text{imp}(\nu_n) &= \sum_{\conf} p(\conf) G_\conf(\nu_n)  \label{eq:gf:imp} \\
&= p(\conf=0) G_{\conf=0}(\nu_n) + p(\conf=1) G_{\conf=1}(\nu_n) \notag \\
&= (1 - n_f) \mathcal{G}_0(\nu_n) + n_f [\mathcal{G}_0^{-1}(\nu_n) - U]^{-1}, \notag
\end{align}
where $n_f$, the average $f$-electron occupation of the impurity, is given by
\begin{align}
 1- n_f = p(\conf=0) = \frac{Z_{\conf=0}}{Z} , \\
 n_f = p(\conf=1) = \frac{Z_{\conf=1}}{Z} ,
\end{align}
see Eq.~(II.8) in Ref.~\onlinecite{Brandt1989}.
Thus, the ``impurity solver'' Eq.~\eqref{eq:gf:imp} is an analytical formula in this case. Note that Eqs.~\eqref{eq:weiss}-\eqref{eq:gf:imp} do not involve spatial or momentum labels, since they refer to a single-site impurity model.
The Dyson equation gives the impurity self-energy $\Sigma_\text{imp}(\nu_n)$ as
\begin{align}
 \Sigma_\text{imp}(\nu_n) = \mathcal{G}_0^{-1}(\nu_n) - G_\text{imp}^{-1}(\nu_n).
 \label{eq:sigma:imp}
\end{align}
DMFT corresponds to using the impurity self-energy as the approximation for the full lattice self-energy
\begin{equation}
  \Sigma(\nu_n, \kv) \approx \Sigma_\text{DMFT}(\nu_n) \equiv \Sigma_\text{imp}(\nu_n)
\end{equation}
giving the DMFT lattice Green's function as,
\begin{align}
 G_\text{DMFT}(\nu_n, \kv) = [i\nu_n + \mu - \epsilon_\mathbf{k} - \Sigma_\text{DMFT}(\nu_n)]^{-1},
 \label{eq:gf:lattice:dmft}
\end{align}
where $\epsilon_\mathbf{k} = -2t (\cos k_x + \cos k_y)$ is the non-interacting dispersion, the Fourier transform of the tight-binding Hamiltonian. Its local part is obtained as the Brillouin zone average,
\begin{align}
 G_\text{DMFT,loc}(\nu_n) = \frac{1}{N_{\kv}} \sum_{\kv} G_\text{DMFT}(\nu_n, \kv).
 \label{eq:gf:local}
\end{align}

\begin{figure}
    \centering
    \includegraphics[width= \columnwidth]{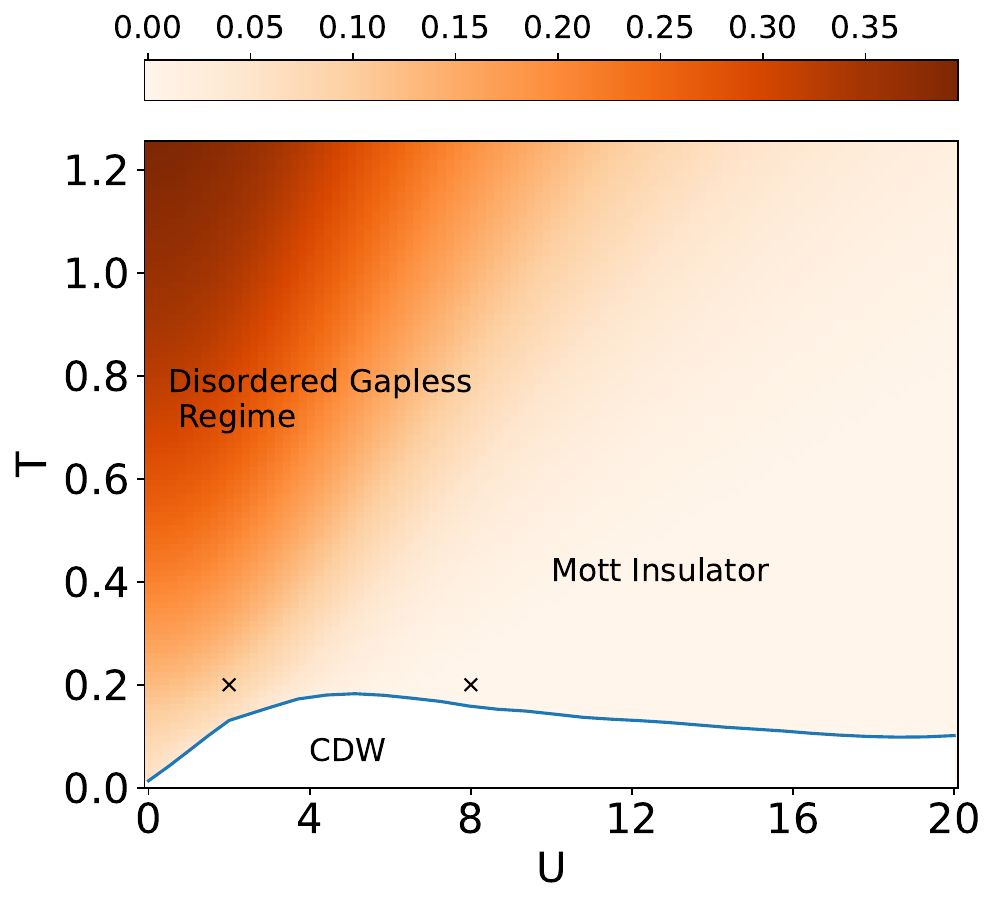}
    \caption{The $T-U$ phase diagram illustrated using the middle point of the local, imaginary time Green's function $\frac{1}{N_\kv} \sum_{\kv} \Im G(\tau = \beta /2, \kv)$ obtained using ladder Dual Fermion for a $16 \times 16$ lattice at half-filling, i.e. $n_c = n_f = 0.5$. These results are consistent with Ref.~\cite{AntipovPRL2014}. The blue line shows an estimate of the critical temperature, $T_c$, from the divergence of $cc$-susceptibility. The two markers highlight the points close to the transition into the CDW phase that we focus on in this paper i.e. $\beta = 5.0$ with $U=2.0$ and $U=8.0$.}
    \label{fig:Gtaubetahalf-DF-l16-muUhalf}
\end{figure} 

\begin{figure*}
    \centering
    \includegraphics[width= \textwidth]{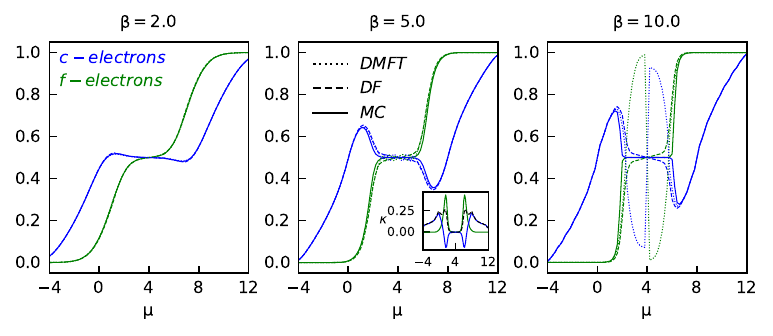}
    \caption{Benchmarking the electron densities $n_c$ (blue) and $n_f$ (green) in DF and DMFT against the numerically exact Monte Carlo results. The densities are shown as a function of the chemical potential $\mu$ with $U=8.0$ and $\beta = 2.0$, $5.0$ and $10.0$, from left to right, respectively. The inset shows compressibilities $\kappa_c$ (blue), $\kappa_f$ (green) and $\kappa = \kappa_c + \kappa_f$ (dashed) as a function of chemical potential from the Monte Carlo simulations. Note that $\kappa$ always remains positive.}
    \label{fig:nc-bCombined-u8-muUhalf}
\end{figure*}

At this point, the Weiss field $\mathcal{G}_0$ or equivalently the hybridization $\Delta_c$ are still free variables. DMFT is based on the self-consistency condition~\cite{GeorgesRMP1996} $G_\text{loc} = G_\text{imp}$, which can be achieved by performing the update  
\begin{align}
 \mathcal{G}_0^{-1}(\nu_n) = G_\text{loc}^{-1}(\nu_n) + \Sigma(\nu_n)
 \label{eq:gf:bare}
\end{align}
so that Eqs.~(\ref{eq:gf:imp}), (\ref{eq:sigma:imp}), (\ref{eq:gf:local}) and (\ref{eq:gf:bare}) form a self-consistency loop that is repeated until convergence is reached. 

While DMFT provides the exact solution for an infinite-dimensional lattice, it is approximate in finite dimensions. By construction, DMFT assumes that the self-energy is local (momentum-independent), which is not true for the exact solution. Diagrammatic extensions of DMFT, such as the ladder Dual Fermion~\cite{RubtsovPRB2008,Hafermann09} approach used here, expand around the DMFT solution. In this way, they incorporate non-local correlations through Feynman diagrams based on impurity vertex functions to address this deficiency. This is especially beneficial close to phase transitions where the non-local correlations are strong. 

Formally, Dual Fermion can be used to obtain the exact self-energy by summing over all diagrams involving all possible reducible impurity vertices (two-particle, three-particle, etc.). It can be noted that the higher-order odd vertices (three-particle, five-particle, etc.) vanish for the Falicov-Kimball model at particle-hole symmetry~\cite{Ribic17}. In practice, summing over all diagrams with all vertices is usually not feasible and a specific subset of vertices and diagrams is chosen. The ladder Dual Fermion used here neglects all higher-order vertices, and for the diagrams involving two-particle vertices it only takes the subset of ladder diagrams. The goal of this paper is to benchmark how appropriate this choice is. 
  
The Dual Fermion perturbation theory~\cite{RubtsovPRB2008,Hafermann09} is based on the dual Green's function $\tilde{G}(\nu_n, \kv)$. Starting from DMFT, the initial guess or bare dual Green's function $\tilde{G}_0(\nu_n, \kv) = G^\text{DMFT}(\nu_n, \kv) - G_\text{imp}(\nu_n)$ is the non-local part of the DMFT Green's function. 
The ladder approximation for the dual fermions is based on a Dyson-like equation for the dual two-particle vertex function $\tilde{\Gamma}$,
\begin{multline}
  \tilde{\Gamma}(\omega, \nu,\nu',\qv) = \gamma(\omega, \nu,\nu')
  - \sum_{\nu'', \kv} \gamma(\omega, \nu,\nu'')
  \\
  \times
  \tilde{G}(\nu'', \kv) \tilde{G}(\nu'' + \omega, \kv + \qv) \, \tilde{\Gamma}(\omega, \nu'',\nu',\qv),
 \label{eq:vertex:dual}
\end{multline}
where $\gamma(\omega, \nu,\nu')$ is the two-particle vertex of the DMFT impurity problem.
For the Falicov-Kimball model, the impurity two-particle Green's function $G^{(2)}$ and $\gamma$ can be obtained in the same way as the single-particle Green's function by summing over the two $f$-electron configurations $\conf=0$ and $\conf=1$, which leads to 
 \cite{Freericks03,RibicPRB2016}
\begin{multline}
  \gamma(\omega, \nu,\nu') =  n_f (1 - n_f) U^2 ( \delta_{\omega 0} - \delta_{\nu \nu'} ) A_{\nu} A_{\nu' + \omega} , 
\end{multline}
where $A_{\nu} = \mathcal{G}_0(\nu) \cdot [\mathcal{G}_0^{-1}(\nu) - U]^{-1} G^{-2}_{\text{imp}}(\nu) $ . The two-particle properties of the Falicov-Kimball impurity model also play a role in the calculation of the conductivity and have been studied in more detail in Refs.~\cite{Janis01,Freericks03,Janis14}.

Note that in the ladder approximation, the vertex $\tilde{\Gamma}(\omega, \nu, \nu', \qv)$ in Eq.~\eqref{eq:vertex:dual} depends only on a single transferred momentum $\qv$ . This vertex is then used to evaluate the dual self-energy by connecting two ends of the vertex with a dual Green's function, 
\begin{align}
 \tilde{\Sigma}(\nu_n, \kv) = \!\!\!\! \sum_{\qv, \omega_m, \nu_{n'}} \tilde{\Gamma}(\omega_m, \nu_n, \nu_{n'}, \qv) \tilde{G}(\nu_{n'} + \omega_m, \kv + \qv)
 \label{eq:sigma:dual}
\end{align}
Then, the Dyson equation is used to obtain the dual Green's function as
\begin{align}
 \tilde{G}^{-1}(\nu_n, \kv) = \tilde{G}_0^{-1}(\nu_n, \kv) - \tilde{\Sigma}(\nu_n, \kv) .
 \label{eq:gf:dual}
\end{align}
Equations~\ref{eq:vertex:dual}, \ref{eq:sigma:dual} and \ref{eq:gf:dual} are solved self-consistently until $\tilde{G}(\nu_n, \kv)$ has converged. Finally, the dual Green's function is transformed to the lattice Green's function $G^\text{DF}(\nu_n, \kv)$ via the relation \cite{RubtsovPRB2008},
\begin{multline}
  G^\text{DF}(\nu_n, \kv) = (\Delta_c(\nu_n) - \epsilon_{\kv})^{-1}
  \\
 + [G_\text{imp}(\nu_n)( \Delta_c(\nu_n) - \epsilon_{\kv})]^{-2} \tilde{G}(\nu_n, \kv) .
\end{multline}
Here, $G_\text{imp}(\nu_n)$ is the Green's function of the impurity model, for example the previously converged DMFT impurity model, and $\Delta_c(\nu_n)$ is the hybridization function of this impurity model. The self-energy $\Sigma(\nu_n, \kv)$ can now be extracted from the single-particle Green's function using the Dyson equation. 

In addition to the single-particle quantities, it is also possible to calculate the susceptibility in Dual Fermion. The dynamical two-particle vertices of the real ($\Gamma$) and dual ($\tilde{\Gamma}$) fermions are related~\cite{Brener08,AntipovPRL2014},
\begin{multline}
  \Gamma(\omega, \nu, \nu', \kv, \kv', \qv) =
  L(\nu,\kv) L(\nu + \omega,\kv+\qv) \tilde{\Gamma}(\omega,\nu,\nu',\qv) \notag 
  \\
  \times
  L(\nu',\kv') L(\nu' + \omega,\kv'+\qv),
\end{multline}
\\
where $L(\nu,\kv) = [1 - \tilde{\Sigma}(\nu, \kv) G_{\text{imp}}(\nu)]^{-1}$. Thereafter, the static $cc$-susceptibility, at the zero bosonic Matsubara frequency, is given as \cite{AntipovPRL2014},
%
\begin{multline}
  \chi_{cc}(\omega_m, \qv) =
  -T \sum_{\nu_n, \kv} G^\text{DF}(\nu_n, \kv)G^\text{DF}(\nu_n + \omega_m, \kv+\qv) 
  \\
  + T \sum_{\nu_n, \kv }\sum_{\nu_{n'},\kv'} G^\text{DF}(\nu_n,\kv) G^\text{DF}(\nu_n + \omega_m,\kv+\qv)
  \\
  \Gamma(\nu_n, \nu_{n'}, \kv, \kv', \qv)
  G^\text{DF}(\nu_{n'},\kv') G^\text{DF}(\nu_{n'} + \omega_m,\kv'+\qv) .
\end{multline}

These methods are implemented in Antipov's DF and DMFT solvers \cite{AntipovDFcode2015}. For more details of our implementation, see Appendix \ref{app:code}.


\section{Results}\label{sec:results}

We start by characterizing the phase diagram of the Falicov-Kimball model at half-filling, following Refs.~\cite{AntipovPRL2014,AntipovPRL2016}. For this purpose, we probe the middle point of the imaginary time Green's function, $G(\tau = \beta /2, \kv)$ which is an approximant for the density of states at the Fermi level and therefore allows us to distinguish between gapless and gapped single-particle spectra. Its local part $\frac{1}{N_k} \sum_\kv G(\tau=\beta/2, \kv)$ is shown as a function of temperature and interaction strength in Fig.~\ref{fig:Gtaubetahalf-DF-l16-muUhalf}.  The results are obtained using ladder Dual Fermion on a $16\times 16$ lattice. The Green's function has converged to the thermodynamic limit at $L=16$, as detailed in Appendix \ref{app:finite-size}. The hopping parameter $t = 1$ sets the energy scales. \\

For temperatures $T \gtrsim 0.2$, we see a crossover from a gapless to a gapped regime as the interaction strength is increased. It should be noted that the absence of a single-particle gap does not mean that the system is a conductor, since disorder in this model leads to an Anderson insulating phase with finite density of states at the Fermi level~\cite{AntipovPRL2016}. Decreasing temperature for a fixed value of interaction strength ($U>0$), we observe a transition to a charge-ordered phase indicated by a diverging $cc$-susceptibility. We should point out that dual fermion shows a diverging susceptibility already in finite size simulations, whereas the exact solution and the MC results will only show a phase transition in the thermodynamic limit~\cite{newman1999monte}. In this sense, Fig.~\ref{fig:Gtaubetahalf-DF-l16-muUhalf} is indicative of the physical regime of the DF solution and not a precise estimate of the phase boundaries of the exact solution.

The temperature $T=0.2$, i.e. $\beta = 5.0$, is close to but still not in the CDW phase, according to Dual Fermion, and is therefore a challenging but feasible case for benchmarking the method. The values $U = 8$ and $U=2$ marked by stars are selected to represent the strongly and weakly correlated regimes. Staying above the CDW phase ensures that we can reliably compare the results from all three methods: DMFT, DF and Monte Carlo. 

\subsection{Electronic Occupation}

Figure~\ref{fig:nc-bCombined-u8-muUhalf} shows the average occupation for the $c$- and $f$-electrons as a function of chemical potential, for three different temperatures and $U=8$. Particle-hole symmetry around half-filling ($\mu=U/2=4$, $n_c=1/2$, $n_f=1/2$) is visible here. In the Monte Carlo data, there is a plateau close to half-filling in both $n_c$ and $n_f$ at all temperatures, which gets more flat as the temperature is lowered. For the $c$-electrons, this plateau is a signal for a gap in the density of states, which is a sign of an insulating phase~\cite{FalicovKimball}.  Here, the insulating state arises due to repulsive interaction between the $c$- and $f$-electrons, which means the system wants to avoid double occupation~\cite{Freericks03}.

The plateau is visible in the small values of the compressibilities $\kappa_c=\partial n_c/\partial \mu$ and $\kappa_f=\partial n_f/\partial \mu$ shown in the inset. Thermodynamic stability requires that the total $\kappa = \partial n/\partial \mu=\kappa_c+\kappa_f$ is positive, which is the case here, but $\kappa_c < 0$ is thermodynamically allowed and occurs for all shown temperatures in a region close to half-filling. The negative $\partial n_c/\partial \mu$ is enhanced at lower temperature. $\partial n_f/\partial \mu = \kappa_f < 0$ is also allowed but does not occur. 

In the regions where $n_f\approx 0$ or $n_f\approx 1$, the $c$-electrons experience a homogeneous potential and there are no correlation effects apart from a shift of the $c$-electron band by $U n_f$. All three methods therefore agree on the occupation in these two regimes. Closer to half-filling, the methods predict different occupations. DMFT performs as well as or even better than Dual Fermion in large regions of parameter space, the main exception is that DMFT underestimates the $c$-occupation and overestimates the $f$-occupation for low temperature and chemical potential just below half-filling. By particle-hole symmetry, the opposite error occurs just above half-filling.  

An impurity-based approach like DMFT can naturally describe local observables such as the occupation numbers, so it is no big surprise that these come out close to the Monte Carlo reference values. The exception is the region at low temperature and close to half-filling where charge-density waves with non-uniform $f$-electron density are starting to be favored. The single-site DMFT used here is based on a uniform density and thus cannot describe this effect. Instead, it starts to favor a uniform polarized solution, where the density is dominated by one of the two electron species.

For comparison, results at $U=2$ are shown in Appendix~\ref{app:standard}. Correlation effects are much weaker in this case, and all three methods give very similar results.

\begin{figure}
    \centering
    \includegraphics[]{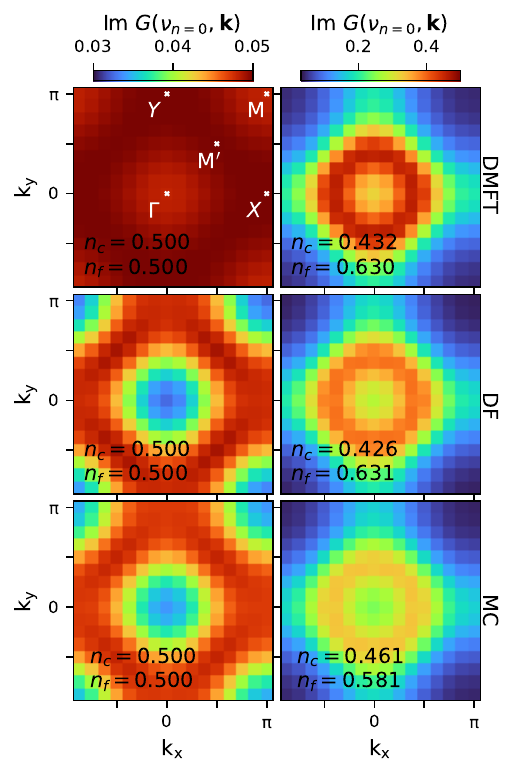}
    \includegraphics[]{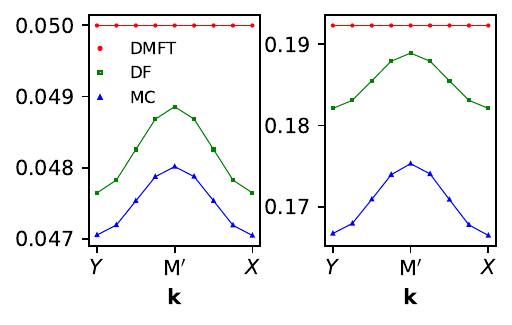}
    \caption{(a) The Green's function at the lowest Matsubara frequency, $\Im G(\nu_{n=0}, \kv)$, in the BZ for $U=8.0$ and $\beta = 5.0$. Results are for a $16 \times 16$ lattice, every pixel corresponds to one of the momenta in the calculation and the $\Gamma$ point is in the bottom left corner. 
    From top to bottom, results for DMFT, DF and MC are shown.
    The left column shows half-filling ($n_c=0.5$), while the right column is at fixed $\mu=6.0$, which leads to slightly different occupations in the three methods, cf. Fig.~\ref{fig:nc-bCombined-u8-muUhalf}. 
    (b) The Green's function at the lowest Matsubara frequency, $\Im G(\nu_{n=0}, \kv)$, plotted along a high symmetry path along the Fermi surface in the BZ.}
    \label{fig:Giw0k-b5.0-u8.0-l16}
\end{figure}

\subsection{Single-particle Green's function}

\begin{figure}
    \centering
    \includegraphics[]{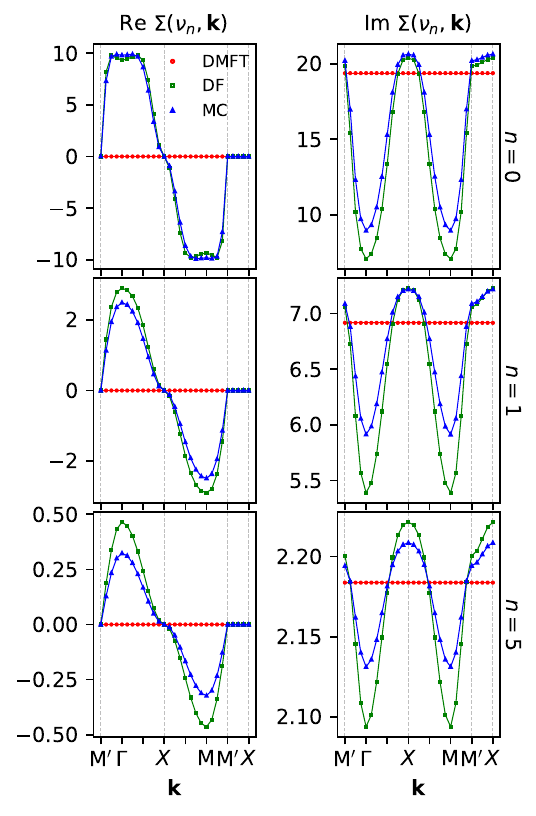}
    \caption{The momentum-dependent self-energy, $\Re \Sigma (\nu_{n}, \kv)$, at different Matsubara frequencies in the BZ for $U=8.0$ and $\beta = 5.0$. Results are for a $16 \times 16$ lattice at half-filling. From top to bottom, results for the zeroth, first and fifth Matsubara frequencies are shown.}
    \label{fig:Sigma-iwk-b5.0-u8.0-l16}
\end{figure}

We continue with a frequency- and momentum-resolved single-particle quantity, namely the Green's function $\Im G(\nu_{n=0},\kv)$ at the lowest Matsubara frequency shown in Fig. \ref{fig:Giw0k-b5.0-u8.0-l16}. The lowest Matsubara frequency gives an indication of the low-energy spectral weight as a function of $\kv$. At half-filling (left column), it is highest on the anti-diagonal from $Y = (0,\pi)$ to $X = (\pi,0)$, where the Fermi surface is. Due to particle-hole symmetry, it is mirror symmetric around this line. The Fermi surface is not perfectly sharp due to finite temperature and the finite lifetime due to correlations. The importance of the latter can be seen by comparing with the result at $U=2.0$, shown in Appendix~\ref{app:standard}, where correlations are not important and the Fermi surface is much sharper. The Green's function varies weakly along the Fermi surface $Y-X$, with a maximum at $M' = (\pi/2,\pi/2)$, see Fig.~\ref{fig:Giw0k-b5.0-u8.0-l16}. This is an indication of the momentum dependence of the self-energy, since the non-interacting dispersion $\epsilon_\kv$ is constant along this path. To get a more quantitative view, Fig.~\ref{fig:Sigma-iwk-b5.0-u8.0-l16} shows the self-energy along the high-symmetry path. By looking at the self-energy instead of the Green's function, the effect of the dispersion is removed and we get a clear view of the non-locality (momentum structure) of the correlations. 

Comparing DMFT and DF to the benchmark MC results in Fig.~\ref{fig:Giw0k-b5.0-u8.0-l16}(c), we see that DF consistently captures the momentum structure of the Green's function, including the maximum at $M'$. Moving to the self-energy, the strongest momentum dependence can be seen in the $M'-\Gamma$ and $X-\Gamma$ directions while the dependence along the Fermi surface is relatively small. Note that particle-hole symmetry ensures that the real part of the self-energy is antisymmetric about the $X$ and $M'$ points on the Fermi surface whereas the imaginary part is symmetric about these points. The imaginary part contributes to the broadening of these electronic states, which is indeed larger at $X$ than at $M'$. DMFT by definition does not have any momentum dependence in the self-energy. The momentum dependence seen here is due to strong correlations, and at lower interaction strengths the self-energy is much more local (see Appendix~\ref{app:standard}). 

\begin{figure}
    \centering
    \includegraphics[]{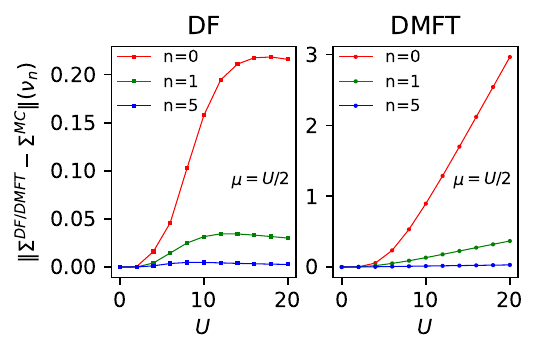}
    \includegraphics[]{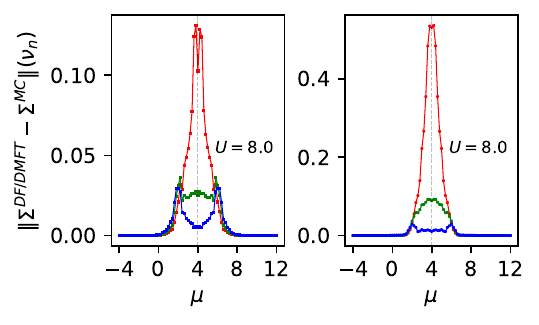}
    \caption{The momentum-normalized deviation in self-energy, $\|\Sigma-\Sigma^\text{MC} \|(\nu_n) =\sqrt{\frac{1}{N_\kv} \sum_\kv \abs{\Sigma(\nu_n,\kv)-\Sigma^\text{MC}(\nu_n,\kv)}^2}$ as a function of interaction strength $U$ (top) and chemical potential $\mu$ (bottom), at different Matsubara frequencies for $\beta = 5.0$. Results are for a $16 \times 16$ lattice with the top panels at half-filling, $\mu = U/2$, and the bottom panels with $U=8.0$.}
    \label{fig:Diff-Sigma-iwk-b5.0-l16-uVar-muVar}
\end{figure}

The systematic overestimation of the momentum dependence of the DF self-energy visible in Fig.~\ref{fig:Giw0k-b5.0-u8.0-l16} can be understood by looking at the Fourier transform to real space, shown in Table \ref{tab:sigma_real_space_n0}. The momentum dependence is dominated by the largest non-local contributions, at $\mathbf{R}=(1,0)$ and $(1,1)$, and dual fermion overestimates the magnitude of both. The self-energy at larger $\mathbf{R}$ is smaller but still clearly non-zero in this parameter regime, with even $\mathbf{R}=(3,2)$ contributing above the 3\% level to the non-local self-energy (note also the different multiplicity of different $\mathbf{R}$-vectors). In cluster extensions of DMFT~\cite{KotliarPRL2001,Maier05}, $\Sigma(\nu_n,\mathbf{R})$ is assumed to be non-zero only for $\mathbf{R}$ within the cluster, so Table~\ref{tab:sigma_real_space_n0} indicates the required cluster size at this point in parameter space.  

\begin{table}[ht]
\centering
\begin{tabular}{r|r|r}
$\mathbf{R}$ & $\mathrm{Im}\, \Sigma$ DF & $\mathrm{Im}\, \Sigma$ MC \\
\hline
$(0,0)$ &  $1.50\times 10^{1}$ & $1.64\times 10^{1}$ 
\\
$(1,1)$ &  -1.99 & -1.70 
\\
$(2,0)$ &  $-6.39\times 10^{-1}$ & $-6.13\times 10^{-1}$ 
\\
$(3,1)$ &  $2.11\times 10^{-1}$ & $1.46\times 10^{-1} $
\\
$(2,2)$ &  $4.21\times 10^{-1}$ & $2.69\times 10^{-1}$ \\
$(3,3)$ &  $-9.77\times 10^{-2}$  & $-4.70\times 10^{-2}$ \\
\hline
  &  $\mathrm{Re}\, \Sigma$ DF & $\mathrm{Re}\, \Sigma$ MC \\
\hline
$(1,0)$ & 3.54 & 3.43  \\
$(3,0)$ & $-6.57\times 10^{-2}$ & $-8.31\times 10^{-2}$  \\
$(2,1)$ & $-7.42\times 10^{-1}$ & $-5.56\times 10^{-1} $  \\
$(3,2)$ & $1.75\times 10^{-1}$  & $9.86\times 10^{-2}$ \\
\end{tabular}
\caption{Real-space self-energy $\Sigma(\nu_{n = 0},\mathbf{R})$ in Dual Fermion and Monte Carlo, for a $16 \times 16$ lattice at half-filling at $U=8.0$, $\beta = 5.0$. Due to particle-hole symmetry, the self-energy is purely imaginary for even $\mathbf{R}$ and purely real for odd $\mathbf{R}$. }
\label{tab:sigma_real_space_n0}
\end{table}

Qualitatively similar dependencies can also be seen for the self-energy at higher Matsubara frequencies. In all cases, DF gets the momentum structure right but overestimates the magnitude of the momentum dependence. The overall magnitude of the momentum dependence compared to the $\kv$-independent part quickly diminishes with increasing frequencies. DMFT thus performs reasonably well for these higher frequencies, which contribute significantly to the evaluation of two-particle observables like the susceptibility which will be discussed in the following subsection. 

The right column of Fig.~\ref{fig:Giw0k-b5.0-u8.0-l16} gives an example away from half-filling at $\mu = 6.0$. This is the regime where $n>1$, $n_f>\frac{1}{2}$ but $n_c<\frac{1}{2}$, i.e., the system is electron-doped overall but the $c$-sector is hole-doped. Thus, the Fermi surface shrinks towards the $\Gamma = (0,0)$ point and becomes more circular. This is seen in the MC benchmark and both approximate methods. Note that this deformation is substantially larger at $U=8.0$ than at $U=2.0$ (see Appendix~\ref{app:standard}), even at similar values of $n_c$. The intensity at the Fermi surface is overestimated in both DF and DMFT, but DF performs better. Note that in the doped case, we still keep $\mu_c = \mu_f = \mu$. \\

\begin{figure}
    \centering
    \includegraphics[width= \columnwidth]{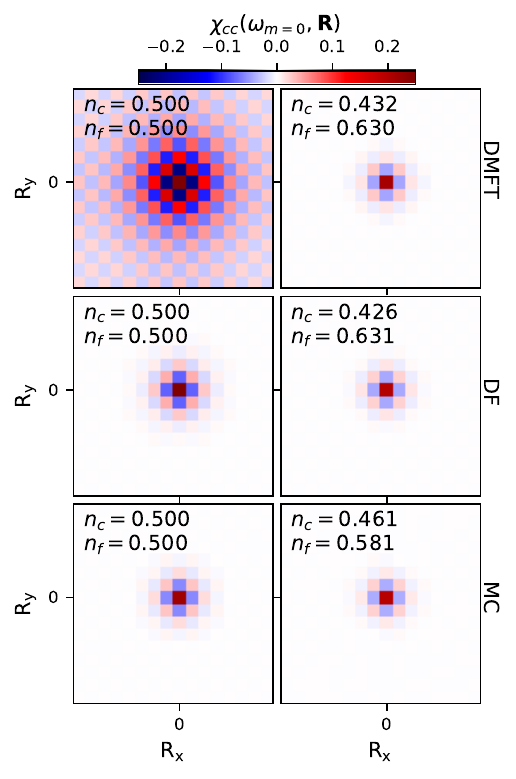}
    \includegraphics[width= \columnwidth]{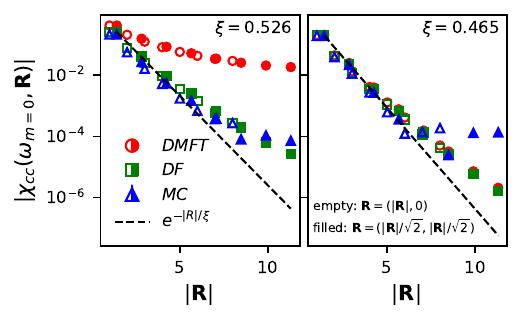}
    \caption{(a) The $cc$-susceptibility at zero frequency, $\chi_{cc}(\omega_{m=0}, \Rv)$ for $U=8.0$ and $\beta = 5.0$. The susceptibility is shown in real space for a $16 \times 16$ lattice, $\Rv=(0,0)$ is at the center of the plot. The left column shows half-filling ($n_c=0.5$), while the right column is at fixed $\mu=6$, which leads to slightly different occupations in the three methods. 
      (b) The decay of the absolute value of $cc$-susceptibility, $|\chi_{cc}(\omega_{m=0}, \Rv) |$ as a function of distance. The dashed line shows an exponential fit $e^{-|\Rv|/\xi}$, where $\xi$ is the correlation length.}
    \label{fig:Chi-iw0R-b5.0-u8.0-l16}
\end{figure}

To quantify more precisely how good the two approximate methods are, their deviation from the MC reference data is shown in Fig.~\ref{fig:Diff-Sigma-iwk-b5.0-l16-uVar-muVar}, as a function of interaction strength and chemical potential. We see that the deviation in the DF results saturates with increasing $U$ and is one order of magnitude smaller than those in the DMFT results (note the different y-axis). As a function of chemical potential, the lowest Matsubara frequency shows the largest deviation at half-filling, whereas the highest shown Matsubara frequency shows a double peak structure. The two peaks mark the inflection points in the variation of the occupation with chemical potential, after which a negative c-electron compressibility is observed ($\kappa_c < 0$). In this region, there are noticeable deviations in electron density between the methods. Since the electron density determines the high-frequency asymptote of the self-energy, it is reasonable that deviations in the density lead to deviations in the self-energy. Note that by particle-hole symmetry, these curves should be symmetric around half-filling, the imperfect symmetry gives an idea of the uncertainty in the MC data.

\subsection{Two-particle Response Function}

We now compare the dynamic cc-susceptibilities, $\chi_{cc}(\omega_{m}, \Rv)$ shown in Fig.~\ref{fig:Chi-iw0R-b5.0-u8.0-l16}, calculated from the three methods. As mentioned previously and shown in Appendix~\ref{app:finite-freq}, it has a dominant contribution at zero Matsubara frequency. The susceptibility is shown in real space as a function of the distance $\Rv$. Its maximum is at $\Rv=0$ and it shows exponential decay, $e^{-|\Rv|/\xi}$, in magnitude, where $\xi$ is the correlation length. At the same time, it shows checkerboard oscillations in sign with respect to $\Rv$. The DF result is similar to the MC benchmark in magnitude and structure. DMFT shows very little spatial decay, which is a sign that DMFT overestimates the tendency towards ordering ($\xi \rightarrow \infty$ as $T\rightarrow T_c$). \\

\begin{figure}
    \centering
    \includegraphics[width= \columnwidth]{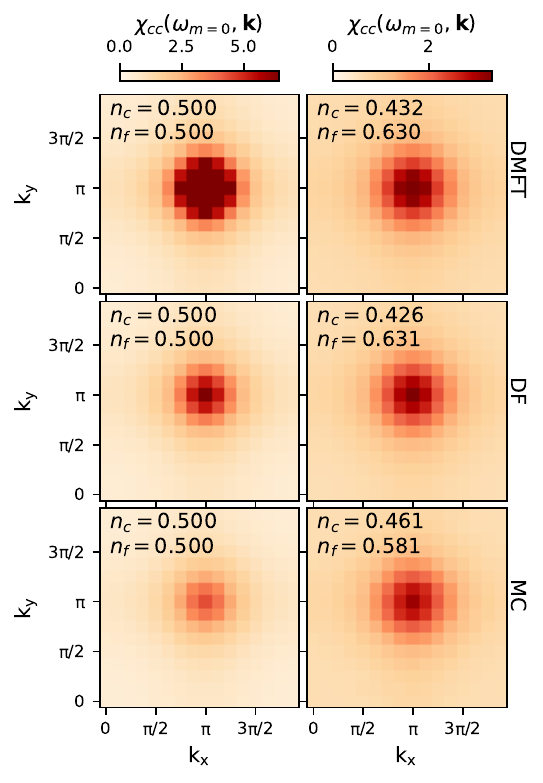}
    \includegraphics[width= \columnwidth]{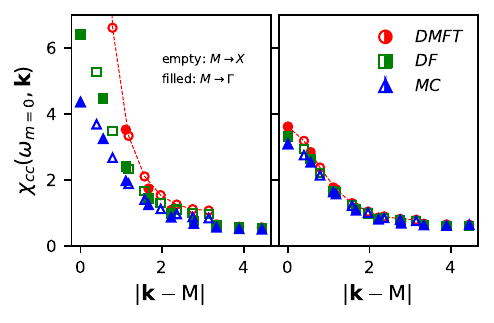}
    \caption{(a) The $cc$-susceptibility at zero frequency, $\chi_{cc}(\omega_{m=0}, \kv)$ in the BZ for $U=8.0$ and $\beta = 5.0$. Results are for a $16 \times 16$ lattice, every pixel corresponds to one of the momenta in the calculation and the $\Gamma$ point is in the bottom left corner. The left column shows half-filling ($n_c=0.5$), while the right column is at fixed $\mu=6$, which leads to slightly different occupations in the three methods, cf. Fig.~\ref{fig:nc-bCombined-u8-muUhalf}. \\
    (b) The decay of the $cc$-susceptibility at zero frequency, $\chi_{cc}(\omega_{m=0}, \kv)$ as a function of distance from the $M$-point in the BZ. The red dashed line shows that the DMFT $cc$-susceptibility values become very large near the $M$-point  at half filling in contrast to the DF and Monte Carlo results.}
    \label{fig:Chi-iw0k-b5.0-u8.0-l16}
\end{figure}

Away from half-filling, the difference between the three methods is much smaller. In both MC and DF, the doped and half-filled model have rather similar susceptibilities for these parameters. Further, the correlation length $\xi$ decreases with doping. At a lower interaction strength, $U = 2.0$ (see Appendix~\ref{app:standard}), all three methods show similar results even at half-filling, due to the reduced tendency towards checkerboard order. \\

Although the differences in real space appear to be relatively small, they show up more clearly in the Fourier transform $\chi_{cc}(\omega_{m}, \kv)$ shown in Fig.~\ref{fig:Chi-iw0k-b5.0-u8.0-l16}. These susceptibilities show a strong peak at $\qv=M$ corresponding to the checkerboard ordering tendencies, and the magnitude of this peak is larger in DF compared to MC for $U=8.0$, $\beta = 5.0$, and half-filling. The DMFT susceptibility here becomes very large at the $M$-point. In the doped case, however, the match in momentum space between DF/DMFT and the MC benchmark is quite good, with DF being slightly better near $\qv=M$. \\

\begin{figure}
    \centering
    \includegraphics[width= 0.98\columnwidth]{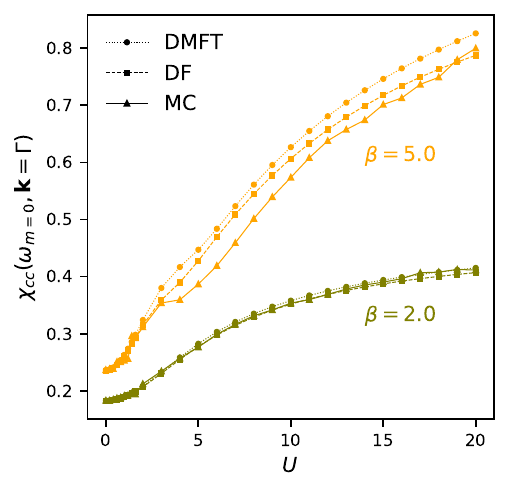}
    \includegraphics[width= 0.98\columnwidth]{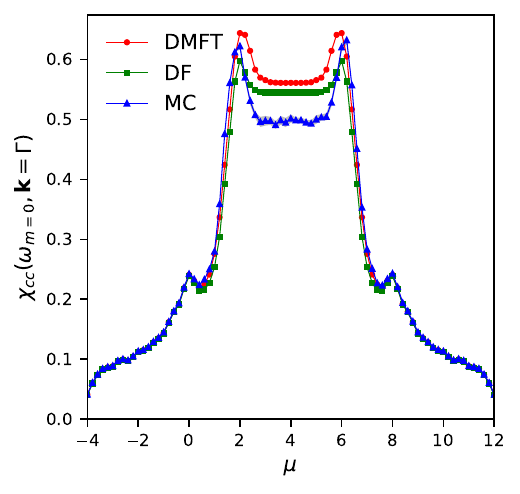}
    \caption{The $cc$-susceptibility at zero frequency, $\chi_{cc}(\omega_{m=0}, \kv = \Gamma)$ at the $\Gamma$-point in the BZ (a) as a function of interaction strength for $\beta = 2.0$ (olive), $5.0$ (orange) at half-filling, i.e. $n_c = 0.5$, and (b) as a function of chemical potential for $U=8.0$, $\beta = 5.0$. DMFT (circle), DF (square) and MC (triangle) results are shown for a $16 \times 16$ lattice. Note that the uncertainty in the Monte Carlo data for $U=8.0$, $\beta = 5.0$ is maximum near half-filling and it falls down rapidly as we move further away. A confidence of $2 \sigma$ is shown in grey. (see details in Appendix \ref{app:errors})}
    \label{fig:ChiCC-DMFT-DF-MC-b2-b5-l16-muUhalf}
\end{figure}

To quantify the agreement in more detail, Fig.~\ref{fig:ChiCC-DMFT-DF-MC-b2-b5-l16-muUhalf} shows the $cc$-susceptibility at the $\Gamma$-point as a function of $U$ for two different values of $\beta$ and as a function of $\mu$ for $U = 8.0$, $\beta = 5.0$. The $cc$-susceptibility at the $\Gamma$-point reflects fluctuations in the total $c$-electron density, or the compressibility $\frac{dn_c}{d\mu_c}$, and is therefore least sensitive to checkerboard fluctuations. The $cc$-susceptibility increases with $U$. This effect is already there at the Hartree level, since an increase in $\mu_c$ will increase $n_c$, which leads to a larger repulsive Hartree potential for the $f$-electrons and therefore decreases in $n_f$, which in turn leads to a reduced Hartree potential for the $c$-electrons and larger $n_c$. Further, the $\mu$-dependence of this $cc$-susceptibility shows deviations only near half-filling. At small $\beta$, all three methods agree, while deviations are visible at larger $\beta$, with a systematic overestimation by both DF and DMFT, but a better performance by DF. \\

\begin{figure}
    \centering
    \includegraphics[width= \columnwidth]{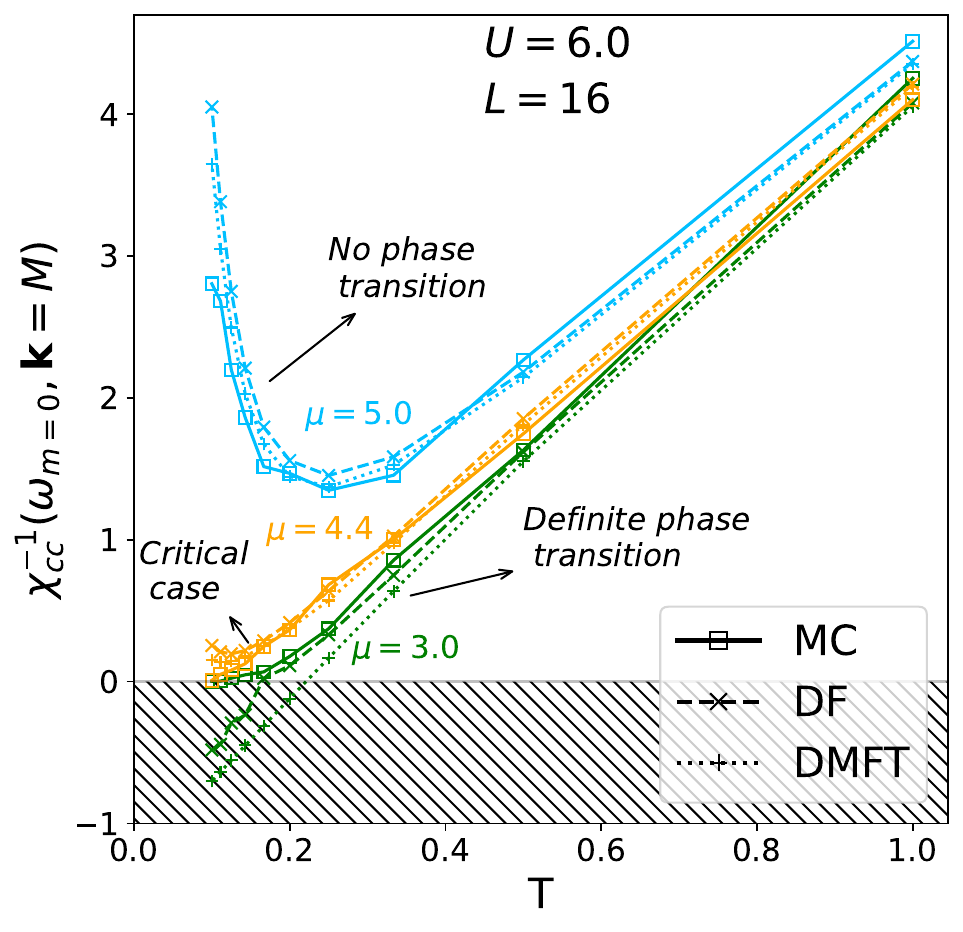}
    \caption{The inverse $cc$-susceptibility, $\chi_{cc}^{-1}(\omega_{m=0}, \kv = M)$, at the $M$-point gives an estimate of the critical temperature $T_c$ at which the phase transition to the checkerboard ordered phase (CDW) takes place. The analysis is done here for $U=6.0$ using DMFT, DF and MC for a $16 \times 16$ lattice. Different colors correspond to different chemical potentials.}
    \label{fig:InvChi-iw0-pipi-u6.0-l16}
\end{figure}

In contrast to the $\Gamma$ point, the $M$ point is most sensitive to checkerboard fluctuations and the susceptibility should diverge in the limit $L\rightarrow \infty$ when there is a CDW transition. For the MC at fixed, finite $L$, the susceptibility remains finite by construction, but $\chi^{-1}$ is eventually very close to zero as the temperature is lowered, see Fig. \ref{fig:InvChi-iw0-pipi-u6.0-l16}. This shows us that DMFT overestimates the critical temperature at half-filling when compared to DF or Monte Carlo. Upon doping, the susceptibility is reduced and the system is no longer close to a phase transition after some critical chemical potential, $\mu_{cr}$. \\

As approximate methods, both DF and DMFT can and do show a divergent susceptibility in a finite system. These methods therefore become unreliable for finite $L$ simulations when they undergo this kind of unphysical transition. Up to that point, though, the DF $cc$-susceptibility is quite close to the MC benchmark and both DMFT and DF perform well in the doped system. \\


\section{Comparison to benchmarks of DF in other models}

It is useful to compare the results of our benchmark with previous benchmarks of (ladder) DF for the two-dimensional Hubbard model. A major physical difference between the two models is the Mermin-Wagner theorem: in the Hubbard model the diagrammatic corrections to DMFT have to push the N\'eel temperature down to zero from its finite DMFT starting point, while the exact solution of the Falicov-Kimball model has a finite N\'eel temperature. Furthermore, controlled reference data is difficult to obtain for the Hubbard model, especially away from half-filling, so that the benchmarks tend to be for individual points in parameter space, whereas our Falicov-Kimball benchmark includes dense scans over chemical potential, and has access to the susceptibility as well as the self-energy.

The large Simons collaboration benchmark~\cite{LeBlanc15} included a comparison of ladder dual fermion's $\Sigma(\nu_n,\kv)$ against the dynamical cluster approximation (DCA) and diagrammatic determinant Monte Carlo (DDMC) at a single point in parameter space ($U/t=8$, $T/t=0.5$, half-filling). It shows that DF is qualitatively reasonable albeit with noticeable quantitative deviations at the lowest Matsubara frequency. 

Iskakov et al.~\cite{Iskakov16} studied the influence of beyond-ladder DF diagrams using a Monte Carlo summation of dual diagrams involving two-particle vertices (DiagMC@DF). At moderate $U$ ($U/t=4$) and half-filling, both ladder DF and DiagMC@DF and DCA reference data agree. At large $U/t$ ($U/t\approx 10$) and half-filling, deviations between the ladder and all diagrams are clearly visible, indicating the limitations of the ladder approach, but no reference DCA data was available. 

Gukelberger et al.~\cite{Gukelberger17} studied ladder DF and DiagMC@DF further, also including several ways of doing self-consistency. They used DDMC and weak-coupling diagrammatic Monte Carlo as their reference methods at half-filling and away from half-filling, respectively, and found that the self-energy in DF is close to the exact result above the DMFT N\'eel temperature, and a breakdown of the DF perturbative series occurs at lower temperature. The parquet DF approach~\cite{Astleithner20}, another way of including non-ladder diagrams involving the two-particle vertex, shows similar deviations from ladder DF in the same parameter regimes~\cite{Krien20}. In our Falicov-Kimball benchmark, we also see worse performance of DF at low temperature, but in our case this is associated with a genuine phase transition in the system.

The Simons collaboration~\cite{LeBlanc15} and the Gukelberger et al.~\cite{Gukelberger17} benchmarks both found that there are substantial deviations in the double occupancy, a scalar observable, even in parameter regimes where the DF self-energy is close to the reference results. This matches our unexpectedly large deviations in the orbital occupancy, another scalar observable. This observation is also somewhat reminiscent of the problems with the scalar $\sum_{\omega,\qv} \chi^\text{RPA}(\omega,\kv)$ in RPA at weak coupling even when $\chi^\text{RPA}(\omega,\kv)$ itself is reasonable~\cite{Vilk97}: errors can and do accumulate when tracing out frequency and momentum.

The ladder DF approach does not take into account any dual self-energy diagrams with three-particle vertices or higher. Previous works have studied the impact of the three-particle vertex by including the ``leading''  diagram that involves such a vertex and a two-particle vertex, and gives a contribution to the local part of the dual self-energy. In the Falicov-Kimball model away from half-filling and at low temperature, this diagram is similar in magnitude to the ladder DF diagrams~\cite{Ribic17}, and therefore potentially relevant. This leading diagram is fully local, so away from half-filling it will contribute to the real part of the local self-energy (see the third panel of Fig.~7 in Ref.~\cite{Ribic17b}, for example, where the Hubbard model is studied) and therefore leads to a change in the electron density. Our benchmark has shown that ladder DF performs poorly for the electron density, and the lack of this diagram involving the three-particle could be the reason. 

Finally, another series of benchmarks has been done for one-dimensional disordered~\cite{Terletska13,Yang14} and interacting-disordered~\cite{Haase17} systems, finding good agreement of the DF Green's function with DCA and a maximum of the deviations at the lowest Matsubara frequency and at intermediate $U$, consistent with the present results.

\section{Conclusion and Outlook}

We have benchmarked DMFT and DF in the Falicov-Kimball model, comparing to numerically exact Monte Carlo,
in terms of the electron densities, the momentum- and frequency-resolved Green's function, self-energy and susceptibility, both at and away from half filling.

Compared to the momentum-independent DMFT self-energy, DF produces the qualitatively correct momentum structure in the self-energy while being quantitatively closer to the exact result than DMFT.
The improvement is especially noticeable at the first fermionic Matsubara frequency and in the strongly correlated regime. Comparing with the exact Monte Carlo results, DF tends to slightly overestimate the magnitude of the momentum dependence.
At very large $U$, the deviation between DF and MC starts to decrease again.

For the susceptibility, DF also performs notably better than DMFT, which substantially overestimates the tendency towards charge order.
Looking specifically at $\chi_{cc}(\omega_m=0,\mathbf{k}=\Gamma)$, the $cc$-compressibility, deviations are enhanced as the temperature is lowered. One notable observation is that DMFT performs slightly better in terms of both the electron density and the $cc$-compressibility in parts of the doped regime. This regime has an average density below half filling, but the $c$-sector is more than half-filled, while the $f$-sector is almost empty. Increasing the chemical potential leads to a sudden change to a situation where both sectors are approximately half-filled. Both DF and DMFT have difficulty describing this sudden change, which is overestimated in DMFT and underestimated in DF. The counterpart of this regime is the Nagaoka ferromagnetism in the Hubbard model, with the difference that ferromagnetism involves spontaneous symmetry breaking while the regime observed here has explicitly broken symmetry which is strongly enhanced by interactions. 

Our benchmarking takes place at temperatures just above where the CDW transition occurs. Note that this transition only appears in the exact solution in the thermodynamic limit, while the finite systems considered here only show strong long-ranged correlations. For both DMFT and DF, we use a single-site impurity, which is problematic close to the ordered phase. Starting from a cluster~\cite{tscheppe2025} or from a non-uniform real-space DMFT solution~\cite{TranPRB2006} with different impurities for the two sublattices could alleviate this problem.  
 
The poor performance of DMFT in the doped square-lattice Hubbard model is one of the main motivations for diagrammatic extensions of DMFT such as DF. Thus, it is useful to reflect on what the current benchmark tells us about the kind of physics that is relevant in that case. One of the main observations in the self-energy of the doped Hubbard model is the dichotomy between the $\Gamma$-X and $\Gamma$-M' directions~\cite{GullPRB2010, WuPRX2018,vsimkovic2024origin}. In the Falicov-Kimball model, at half-filling, we similarly see a noticeable difference between the two directions. This spatial contribution to the self-energy is driven by the large peak in the susceptibility at the M point, which DF is able to capture qualitatively.

Finally, in addition to benchmarking by directly comparing results of different methods, the numerically exact Monte Carlo results also allow us to directly test some of the assumptions in approximate methods. For example, within DMFT the main assumption is that $\Sigma$ is independent of $\kv$, which is clearly not true in the benchmark data. As another example, within the ladder dual fermion approach for the Falicov-Kimball model, the $cc$-susceptibility at $\omega>0$ has to be zero, and the MC results show that this is a very good approximation except at very small $U$ or in the $n_c\approx 0.5$, $n_f\approx 0$ regime. 
 
We consider the promising benchmark results of DF on the Falicov-Kimball model a strong argument for extending the application of DF to \textit{ab initio} derived multi-orbital Hubbard models for real materials. The recent success of dual TRILEX \cite{ts6y-zb6m, j6bj-gz7j, PhysRevLett.132.236504} on this class of problems is also encouraging, since it contains subset of the DF vertex contributions (i.e.\ the triangular vertex part). \\


\acknowledgments

We thank Andrey Antipov for useful discussions and for help with the codes.

A.M. and E.vL. acknowledge support by the Swedish Research Council, (Vetenskapsrådet, VR) under grant 2022-03090 and by eSSENCE, a strategic research area for e-Science, grant number eSSENCE@LU 9:1.
H.U.R.S acknowledges financial support from the Swedish Research Council (Vetenskapsrådet, VR) grant number 2024-04652 and funding from the European Research Council (ERC) under the European Union’s Horizon 2020 research and innovation programme (grant agreement No.\ 854843-FASTCORR).
We used the TRIQS library~\cite{TRIQS} for post-processing data.
The computations were enabled by resources provided by LUNARC, the Centre for Scientific and Technical Computing at Lund University through the projects LU 2025/2-47 and LU 2025/17-18.
 

\bibliography{references}

\break 

\appendix 

\section{Finite Matsubara frequency dependence}
\label{app:finite-freq}

\begin{figure}[H]
    \centering
    \includegraphics[width= \columnwidth]{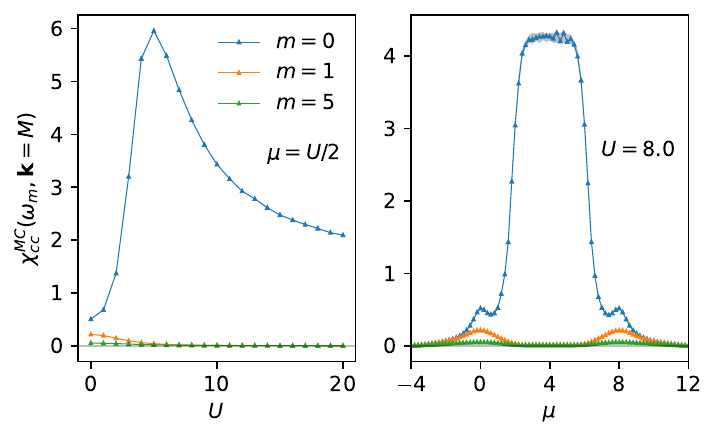}
    \caption{The $cc$-susceptibility at the $M$ point, $\chi_{cc}(\omega_{m}, \kv=M)$ (left) as a function of interaction strength $U$ for $\beta = 5.0$ at half-filling and (right) as a function of chemical potential $\mu$ for $U=8.0$ and $\beta = 5.0$. Results are for a $16 \times 16$ lattice from Monte Carlo simulations. The zeroth, first and fifth bosonic Matsubara frequency contributions are shown. The uncertainty in the Monte Carlo data is shown in grey with a $2 \sigma$-confidence for $U=8.0$, $\beta = 5.0$ and it is maximum near half-filling. See Appendix \ref{app:errors} for details of error analysis.}
    \label{fig:Chi-iwkM-b5.0-l16}
\end{figure}

We compare the Matsubara frequency dependence of the $cc$-susceptibility at the M-point as this is the point most affected by the checkerboard density-density correlations and the susceptibility is the strongest at this point. Figure~\ref{fig:Chi-iwkM-b5.0-l16} shows that the zeroth bosonic Matsubara frequency has the largest magnitude, especially at larger $U$ and close to half-filling. These results ensure that comparing only the most dominant zero bosonic Matsubara frequency contribution is enough for the critical cases that we have focused on in the main text. A similar decoupling of the static ($m=0$) and dynamic ($m>0$) checkerboard susceptibility has previously been observed in the infinite-dimensional limit of the Falicov-Kimball model~\cite{Freericks00}.

\section{Finite size dependence}
\label{app:finite-size}

\begin{figure}[H]
    \centering
    \includegraphics[width= \columnwidth]{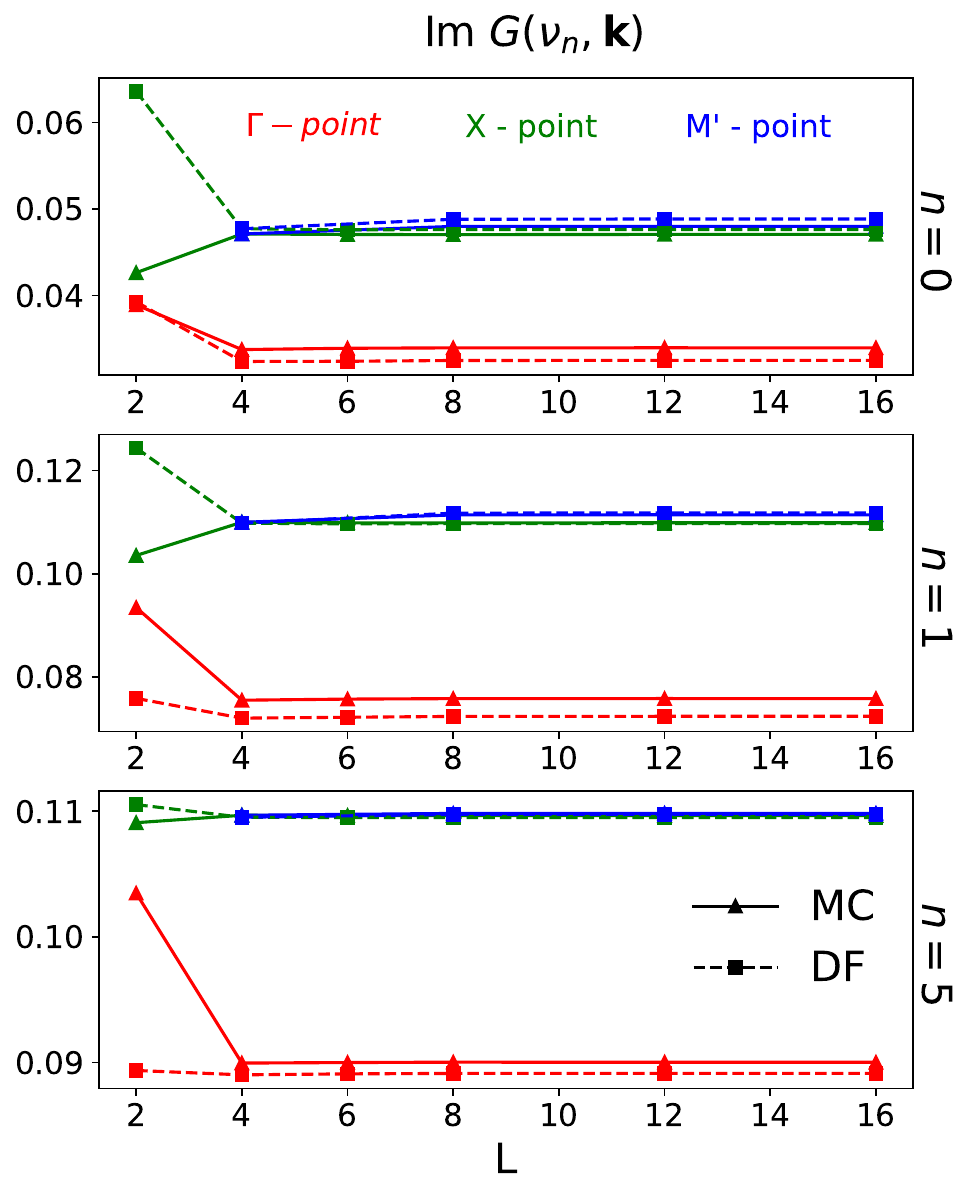}
    \caption{The Green's function, $\Im G(\nu_{n}, \kv)$, at different high symmetry points in the BZ for $U=8.0$ and $\beta = 5.0$ as a function of length L, at half-filling. Results are for a $L \times L$ lattice from Monte Carlo and Dual Fermion simulations. The zeroth (top), first (middle) and fifth (bottom) fermionic Matsubara frequency contributions are shown.}
    \label{fig:G-iwk-b5.0-u8.0}
\end{figure}

The simulations are based on a finite $L\times L$ square lattice, with $L=16$ in the main text. Figure~\ref{fig:G-iwk-b5.0-u8.0} shows how the Green's function, $\Im G(\nu_{n}, \kv)$ depends on $L$. The data is for $U=8.0$, $\beta = 5.0$ and half-filling, which is our main benchmark regime, and data is shown for both MC and Dual Fermion. We see that the values of the imaginary part of the Green's function converge quickly to the thermodynamic limit for $L \ge 4$. This validates the choice of $L=16$ in the main text.

\section{Implementation Details}
\label{app:code}

For the Monte Carlo calculations, we use Antipov's MC solver for the Falicov-Kimball model \cite{AntipovPRL2016} to generate a sampled set of $f$-electron configurations and perform all further MC analysis starting from these configurations. A notable aspect of this MC solver is that it uses Chebyshev polynomials to calculate $\Tr_c \exp(-\beta H_c^\conf)$ instead of performing diagonalization of the tight-binding $c$-electron Hamiltonian. The processing of the configurations is implemented using the Toolbox for Interacting Quantum Systems (TRIQS) \cite{TRIQS} and the Two Particle Response Function toolbox (TPRF) \cite{TPRF}. To compress the Green's functions, we use the imaginary-frequency Discrete Lehmann representation (DLR) \cite{KayePRB2022, KayeCPC2022, KayeJOSS2024} to evaluate Eq.~\eqref{eq:gf:averaging} on a restricted, well-chosen set of Matsubara frequencies from which the entire Matsubara Green's function can be reconstructed. For the calculation of $G^{(2)}$, bubbles of these single-particle Green's functions are evaluated for each configuration in $\conf$, disorder-averaged and then stored in the bosonic Green's function object of TPRF. \\

For the DMFT and Dual Fermion calculations, again Antipov's solvers for the Falicov Kimball model \cite{AntipovPRL2014} are used to evaluate the Green's function and the static susceptibility. The Dual Fermion procedure has two nested self-consistency loops. The outer loop involves solving the impurity problem to obtain the impurity Green's function and impurity vertices. This is only done once in our implementation, that essentially means that $n_f$ is only updated once in Dual Fermion. The inner loop involves computing the dual self-energy and updating the dual Green's function. This is performed self-consistently until convergence and then the lattice Green's function is extracted from these converged dual observables. The data is then finally handled using the Toolbox for Interacting Quantum Systems (TRIQS) \cite{TRIQS} and the Two Particle Response Function toolbox (TPRF) \cite{TPRF}. 

\section{Monte Carlo error analysis}
\label{app:errors}

The uncertainty in the Monte Carlo data is evaluated using binning analysis. We divide the total number of measurements, $\{ A_1 A_2, .., A_N \}$ into $M$ equally sized bins of size $b = N/M$, where $\{A_i\}$ can still be correlated. \\

We then evaluate the average for each bin where the bin average of the $k^{th}$ bin is given by,

\begin{equation}
    B_k = \frac{1}{b} \sum_{i=(k-1)b + 1}^{kb} A_i \ ,
\end{equation}

for $k = 1,2, .., M$ .

The standard deviation of these bin averages is

\begin{equation}
    \sigma_M  = \sqrt{\frac{1}{M-1} \sum_{k=1}^M {(B_k - \bar{A})}^2} , 
\end{equation}

where $\bar{A} = \frac{1}{M} \sum_{k=1}^M B_k$ is the overall average of all $N$ measurements. \\

When the bin size $b$ is large enough, the standard deviation $\sigma_M$ corresponds to an empirical estimate of the Monte Carlo error. For the error estimates shown in Fig.~\ref{fig:ChiCC-DMFT-DF-MC-b2-b5-l16-muUhalf} and Fig.~\ref{fig:Chi-iwkM-b5.0-l16}, we have collected a total of $N = 550000$ measurements and divided them into $M = 11$ equally sized bins, each with $b = 50000$ measurements.

\section{Benchmarking for three-dimensional Falicov-Kimball model}
\label{app:3d}

We focus on the two-dimensional square lattice in the paper because non-local fluctuations are strong in low-dimensional systems~\cite{Metzner89}, making it a more stringent benchmark for testing beyond-DMFT diagrammatic corrections, like the Dual Fermion. Here, we present benchmarks for the three-dimensional Falicov-Kimball model below along the same lines as the two-dimensional benchmarks in the main text. Note that the scales of $U$ and $T$ are different in two-dimensional and three-dimensional systems even after rescaling the hopping~\cite{ZondaSSC2009}, so that it is not possible to make a one-to-one comparison at fixed $U$ and $T$: for the three-dimensional system, $U=8$ and $\beta=5$ is very close to the DF phase transition~\cite{AntipovPRL2014}.

\begin{figure}[H]
    \centering
    \includegraphics[width= \columnwidth]{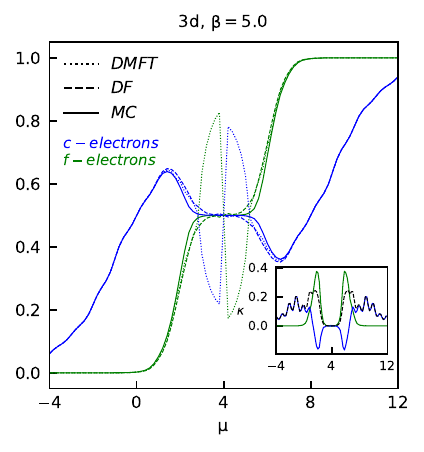}
    \caption{Benchmarking the electron densities $n_c$ and $n_f$ in DF and DMFT against the numerically exact Monte Carlo results for the three-dimensional Falicov-Kimball model with $L=6$. The densities are shown as a function of the chemical potential $\mu$, for $U=8.0$ and $\beta = 5.0$.  The inset shows compressibilities $\kappa_c$ (blue), $\kappa_f$ (green) and $\kappa = \kappa_c + \kappa_f$ (dashed) as a function of chemical potential from the Monte Carlo simulations. Note that $\kappa$ always remains positive.}
    \label{fig:nc-bCombined-u8-muUhalf-l6}
\end{figure}

\begin{figure}[H]
    \centering
    \includegraphics[]{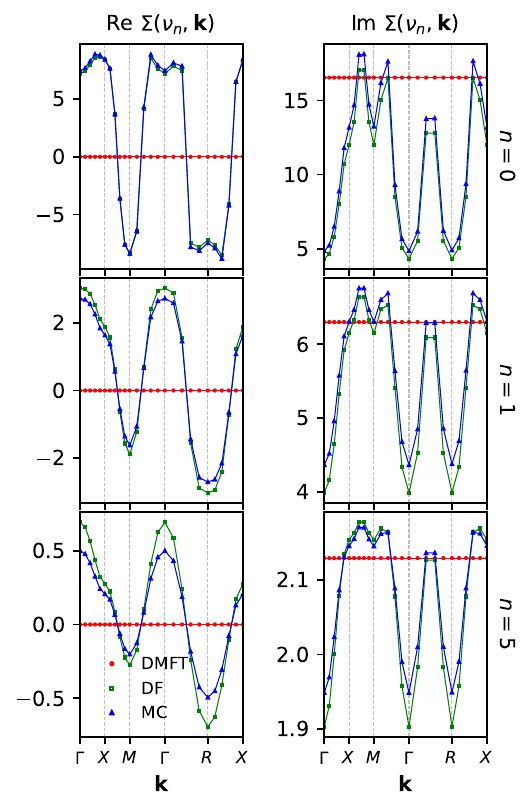}
    \caption{The momentum-dependent self-energy, $\Re \Sigma (\nu_{n}, \kv)$, at different Matsubara frequencies in the BZ for $U=8.0$ and $\beta = 5.0$. Results are for a $10 \times 10 \times 10$ cubic lattice at half-filling. We define the $R = (\pi , \pi , \pi)$ point. From top to bottom, results for the zeroth, first and fifth Matsubara frequencies are shown.}
    \label{fig:Sigma-iwk-b5.0-u8.0-l6}
\end{figure}

The electron densities in the three-dimensional case, see Fig.~\ref{fig:nc-bCombined-u8-muUhalf-l6}, look very similar to the two-dimensional Falicov-Kimball model but the DMFT results fail to describe the non-uniform charge densities even at $\beta = 5.0$.
For the self-energies in the three-dimensional case (see Fig.~\ref{fig:Sigma-iwk-b5.0-u8.0-l6}), we see that the Dual Fermion results are closer to the Monte Carlo benchmark than in the two-dimensional case even though these parameters are very close to the phase transition.   

\section{Benchmarking for low $U$ and moderate $T$}
\label{app:standard}

\begin{figure}
    \centering
    \includegraphics[]{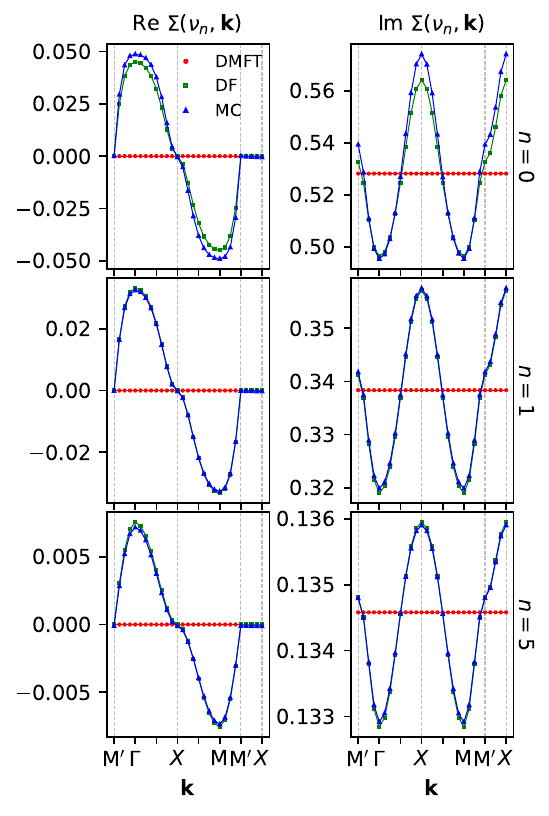}
    \caption{The momentum-dependent self-energy, $\Re \Sigma (\nu_{n}, \kv)$, at different Matsubara frequencies in the BZ for $U=2.0$ and $\beta = 5.0$. Results are for a $16 \times 16$ lattice at half-filling. From top to bottom, results for the zeroth, first and fifth Matsubara frequencies are shown.}
    \label{fig:Sigma-iwk-b5.0-u2.0-l16}
\end{figure}

As has been mentioned in the main text, we show that both DMFT and DF results agree very well with the Monte Carlo results for $U=2.0$ as the dependence of the self-energy on momentum, see Fig.~\ref{fig:Sigma-iwk-b5.0-u2.0-l16}, is very weak in this low correlation regime. \\

\begin{figure*}
    \centering
    \includegraphics[width= \textwidth]{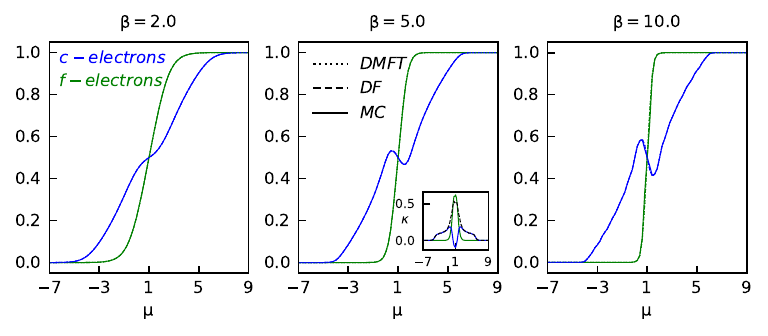}
    \caption{Benchmarking the electron densities $n_c$ and $n_f$ in DF and DMFT against the numerically exact Monte Carlo results. The densities are shown as a function of the chemical potential $\mu$, for a $16 \times 16$ lattice, for $U=2.0$ and $\beta = 2.0$, $5.0$ and $10.0$.  The inset shows compressibilities $\kappa_c$ (blue), $\kappa_f$ (green) and $\kappa = \kappa_c + \kappa_f$ (dashed) as a function of chemical potential from the Monte Carlo simulations. Note that $\kappa$ always remains positive.}
    \label{fig:nc-bCombined-u2-muUhalf}
\end{figure*}

At this interaction strength, the occupation of $f$-electrons is monotonic in $\mu$, so the compressibility $\kappa_f$ is positive, regardless of the temperature or doping. The $c$-electrons still show some negative compressibility in its occupation close to half filling at $\beta=5.0$ and $\beta=10.0$. 

\begin{figure}
    \centering
    \includegraphics[width= \columnwidth]{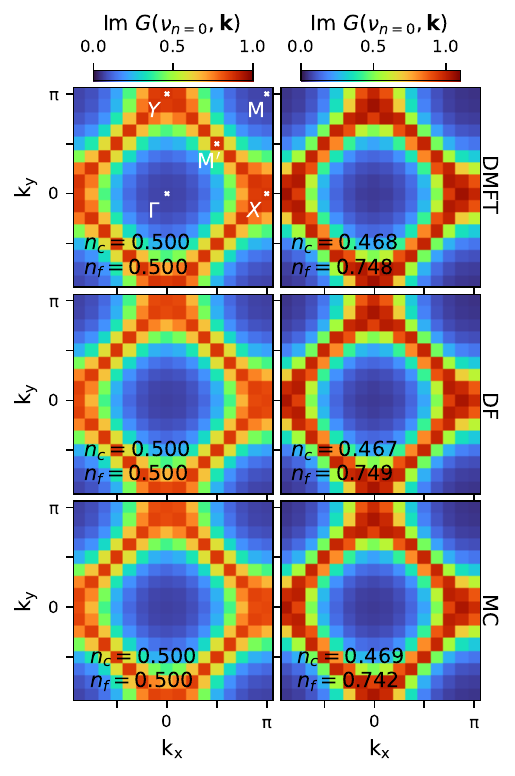}
    \caption{The Green's function at the lowest Matsubara frequency, $\Im G(\nu_{n=0}, \kv)$ in the BZ for $U=2.0$ and $\beta = 5.0$. Results are for a $16 \times 16$ lattice, every pixel corresponds to one of the momenta in the calculation and the $\Gamma$ point is in the bottom left corner. 
    The left column shows half-filling ($n_c=0.5$), while the right column is at fixed $\mu=1.4$, which leads to slightly different occupations in the three methods, cf. Fig.~\ref{fig:nc-bCombined-u2-muUhalf}.}
    \label{fig:Giw0k-b5.0-u2.0-l16}
\end{figure}

The Green's function at low Matsubara frequency, $\Im G(\nu_{n=0}, \kv)$, does not acquire a circular shape around the $\Gamma$-point even for the doped case, see Fig.~\ref{fig:Giw0k-b5.0-u2.0-l16}.

\begin{figure}
    \centering
    \includegraphics[width= \columnwidth]{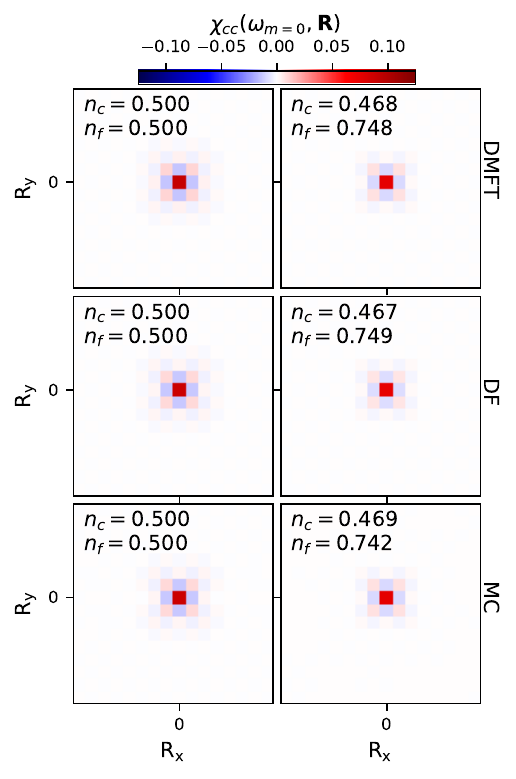}
    \caption{The $cc$-susceptibility at zero frequency, $\chi_{cc}(\omega_{m=0}, \Rv)$ in the real space for $U=2.0$ and $\beta = 5.0$. Results are for a $16 \times 16$ lattice, every pixel corresponds to one of the lattice points. The left column shows half-filling ($n_c=0.5$), while the right column is at fixed $\mu=1.4$, which leads to slightly different occupations in the three methods, cf. Fig.~\ref{fig:nc-bCombined-u2-muUhalf}.}
    \label{fig:Chi-iw0R-b5.0-u2.0-l16}
\end{figure}

\begin{figure}
    \centering
    \includegraphics[width= \columnwidth]{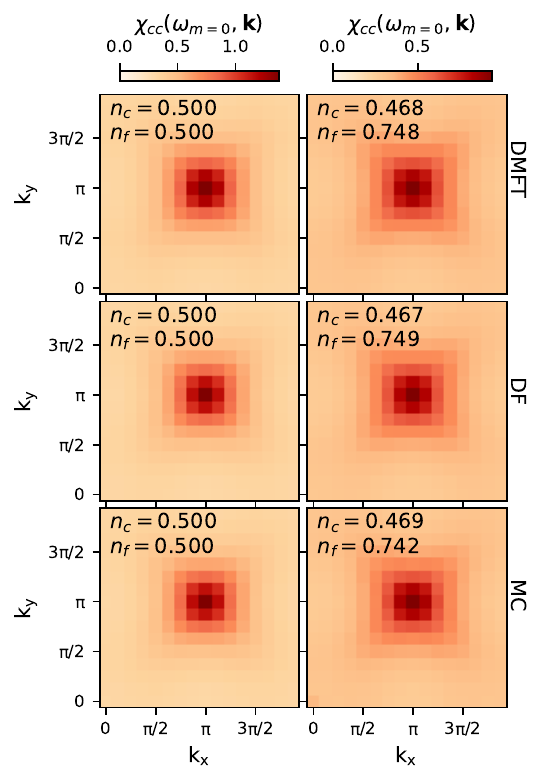}
    \caption{The $cc$-susceptibility at zero frequency, $\chi_{cc}(\omega_{m=0}, \kv)$ in the BZ for $U=2.0$ and $\beta = 5.0$. Results are for a $16 \times 16$ lattice, every pixel corresponds to one of the momenta in the calculation and the $\Gamma$-point is in the bottom left corner. The left column shows half-filling ($n_c=0.5$), while the right column is at fixed $\mu=1.4$, which leads to slightly different occupations in the three methods, cf. Fig.~\ref{fig:nc-bCombined-u2-muUhalf}.}
    \label{fig:Chi-iw0k-b5.0-u2.0-l16}
\end{figure}

As shown in Figs.~\ref{fig:Chi-iw0R-b5.0-u2.0-l16} and \ref{fig:Chi-iw0k-b5.0-u2.0-l16}, the magnitude of the $cc$-susceptibility is significantly lower in this weakly correlated regime as we are further away from the charge ordered phase in the $T-U$ phase diagram.

\end{document}